\documentclass[a1paper, usenatbib]{mnras}

\usepackage[T1]{fontenc}
\usepackage[latin9]{inputenc}
\usepackage{amsmath}
\usepackage{amsfonts}
\usepackage{amssymb}
\usepackage{xcolor}           
\usepackage{graphicx}
\usepackage{natbib}

\title[Low surface brightness galaxies in EAGLE]{Massive Low Surface Brightness Galaxies in the EAGLE Simulation}
\author[A. Kulier et al.]{Andrea Kulier$^{1, 2}$\thanks{E-mail: \href{mailto:akulier@astro.puc.cl}{akulier@astro.puc.cl}}, Gaspar Galaz$^{1}$, 
Nelson D. Padilla$^{1}$, James W. Trayford$^{2}$\\
$^{1}$Instituto de Astrof\'{i}sica, Pontificia Universidad Cat\'{o}lica de Chile, Av. Vicu\~{n}a Mackenna 4860, Santiago, Chile\\
$^{2}$Leiden Observatory, Leiden University, P.O. Box 9513, NL-2300 RA Leiden, The Netherlands
}

\begin{document}

\date{\today}

\pagerange{\pageref{firstpage}--\pageref{lastpage}} \pubyear{2019}

\maketitle

\label{firstpage}

\begin{abstract}
We investigate the formation and properties of low surface brightness galaxies (LSBGs)
with $M_{*} > 10^{9.5} \mathrm{M_{\odot}}$ in the EAGLE hydrodynamical cosmological simulation.
Galaxy surface brightness depends on a combination of stellar mass surface density
and mass-to-light ratio ($M/L$), such that low surface brightness is strongly correlated with both 
galaxy angular momentum (low surface density) and low specific star formation rate (high $M/L$).
This drives most of the other observed correlations between surface
brightness and galaxy properties, such as the fact that most LSBGs have low metallicity.
We find that LSBGs are more isolated than high surface brightness galaxies (HSBGs), in agreement
with observations, but that this trend is driven entirely by the fact that LSBGs are unlikely to
be close-in satellites. The majority of LSBGs are consistent with a formation scenario
in which the galaxies with the highest angular momentum are those that formed most of their stars recently from
a gas reservoir co-rotating with a high-spin dark matter halo. 
However, the most extended LSBG disks in EAGLE, which are comparable in size to observed giant
LSBGs, are built up via mergers. These galaxies are found to inhabit dark matter halos with a higher 
spin in their inner regions ($<0.1r_{200c}$), even when excluding the effects of baryonic
physics by considering matching halos from a dark matter only simulation with identical initial conditions.
\end{abstract}


\section{Introduction}
\label{intro}

Low surface brightness galaxies (LSBGs) are galaxies whose disks are
at least one magnitude fainter than the typical sky brightness \citep{impeybothun}. 
The exact surface brightness value delimiting this category of objects varies
throughout the observational literature, but a common definition is
galaxies having a central disk surface brightness 
fainter than 22 to 23 mag/asec$^{2}$ in the B band \citep{impey}.
These galaxies are challenging to observe due to their faintness, but
surveys of their population in the local Universe suggest that they constitute 
most of the total number density of low-mass galaxies \citep{dalcanton}, and 
$\sim10\%$ of the cosmic baryon budget \citep{minchin2004}. At higher masses,
so-called `giant'
LSBGs like Malin 1 \citep{bothun1987} and UGC1382 \citep{hagen2016}
have the largest known disks in the Universe,
whose extreme sizes potentially challenge our understanding of galaxy assembly \citep{galaz2015, boissier2016}.
Thus LSBGs are an important but still poorly understood component of the overall galaxy population.

A number of authors have found that the properties of LSBGs are statistically different
from those of high surface brightness galaxies (HSBGs), likely
indicating a different formation or evolution scenario for LSBGs. LSBGs have been
found to live in low-density environments and to 
have a deficit of nearby neighbors \citep{bothun1993, mo1994, rosenbaum2004, rosenbaum2009,
galaz2011, du2015}, implying that isolation may be necessary
for either LSBG formation or survival.
LSBGs have also been measured to have low star formation rates
\citep{vanderhulst1993, vandenhoek2000} and low metallicities
\citep{mcgaugh, deblok1998, burkholder2001}.
Some works find that the star formation rates of LSBGs are not unusual
relative to their stellar mass (\citealt{galaz2011}, and references therein), but that their
richness in H\textsc{i} implies low star-formation efficiency \citep{wyder2009, leisman2017}.
Additionally, LSBGs have been reported to be highly dark matter
dominated \citep{deblok1996, pickering, lelli2010}, which is thought
to prevent bar formation that could destroy their diffuse disks 
(\citealt{mayer2004}; see however \citealt{galaz2006}).

Complicating the theoretical interpretation 
of these observations is the possibility that there
are distinct sub-populations of LSBGs. So-called `ultra-diffuse galaxies' (UDGs)
have luminosities and stellar masses
typical of dwarf galaxies, but effective radii $r_{\mathrm{eff}} \gtrsim 1.5$ kpc \citep{vandokkum2,
roman2017}. Unlike more massive LSBGs, observed UDGs
tend to be red, dispersion-dominated, and located within clusters 
\citep{sandage, vandokkum2, vandokkum2015, koda2015, munoz2015, mihos2015, vandokkum2019}.
However, this is at least partly due to selection effects, as 
UDG searches have typically focused on cluster regions. 
UDGs have also been discovered in less dense environments \citep{merritt2016, vdb2017, papastergis2017},
where they are more likely to be blue \citep{leisman2017, zaritsky2019, prole2019}
and rotation-dominated \citep{mancera2019},
and where some have suggested they may be even more numerous
than in clusters \citep{roman2017b, roman2017, mancera2018}. 

Giant LSBGs are also known to
have peculiar features relative to their smaller counterparts.
Specifically, they have bright nuclei that resemble
galaxies of ordinary size, surrounded
by an extremely extended faint disk \citep{barth2007, hagen2016}.

Several different theoretical formation scenarios for low surface brightness galaxies
exist in the literature. For UDGs, hydrodynamical zoom-in simulations 
have suggested that they have very bursty star formation histories, which lead to
episodes of strong feedback that expel their gas
and cause expansion of their stellar orbits \citep{dicintio2017, chan2018}. 
Another proposed UDG formation channel is the expansion of dwarf
galaxies with cored dark matter halos via tidal stripping and heating 
\citep{carleton2019}.

One idea that has been put forth to explain both UDGs and higher-mass
LSBGs is that these galaxies 
constitute the tail of the spin distribution of the galaxy population, forming in the most
high-spin dark matter halos of a given mass \citep{dalcantonb,jimenez1998,amorisco,rong2017}.
However, for giant LSBGs, this would require
extreme halo spins possibly inconsistent with $\Lambda$CDM predictions \citep{boissier2016, zhu}.
For these galaxies, it has been suggested that they build up their 
large outer disks via tidal disruption and accretion of small, 
gas-rich satellites \citep{penarrubia2006, hagen2016}.
Other proposed scenarios for giant LSBG formation include evolution 
from ring galaxies \citep{mapelli2008}, formation
in rare dark matter peaks within voids \citep{hoffman1992}, and
evolution from HSBGs via disk instabilities \citep{noguchi2001}.
 
One newly-opened avenue to exploring the formation and evolution
of LSBGs is to utilize large-scale hydrodynamical cosmological simulations. Historically,
limitations in computational power restricted the possibility of simulating large
samples of well-resolved galaxies in their cosmological environment. However, recent years
have seen the development of large simulations that produce statistically significant
samples of galaxies with realistic properties, 
including EAGLE \citep{eagleschaye, eaglecrain}, Horizon-AGN \citep{horizondubois, horizonkaviraj}, 
Illustris \citep{illustris1, illustris2}, and IllustrisTNG \citep{tng}.

\citet{martin} studied LSBGs in the Horizon-AGN simulation, focusing on galaxies
in the stellar mass range $10^{8} - 10^{10} \mathrm{M_{\odot}}$. They found that ultra-diffuse galaxies (UDGs), 
defined to have $r$-band effective surface brightness $\langle \mu_{e} \rangle > 24.5$ mag/asec$^{2}$,
 and `classical' LSBGs, with $23 < \langle \mu_{e} \rangle < 24.5$ mag/asec$^{2}$, are
generally distinct in their properties. UDGs in Horizon-AGN have stellar masses
$M_{*} < 10^{9} \mathrm{M_{\odot}}$, tend to be very gas-poor, and are 
typically found in dense environments
such as clusters. Dynamical heating via numerous tidal interactions
appears to play an essential role in their formation.
`Classical' LSBGs have properties and evolutionary 
histories more similar to those of HSBGs,
although they formed most of their stars earlier and have lower present-day star formation.

Using hydrodynamical zoom-in simulations of twelve 
galaxies from the NIHAO suite \citep{nihao}
with $10^{9.5} < M_{*}/\mathrm{M_{\odot}} < 10^{10}$, 
\citet{dicintio2019} found that LSBGs
tend to form in dark matter halos with higher spins than
HSBGs, and that surface brightness does not correlate
significantly with any other halo parameters. Galaxies 
and halos with higher angular momentum were found to have experienced
more aligned rather than misaligned mergers. Notably, the authors
also identify distinct formation mechanisms for LSBGs 
with $M_{*} > 10^{9.5} \mathrm{M_{\odot}}$ and those with lower masses
that are UDG-like \citep{dicintio2017}.

Focusing on higher masses, \citet{zhu} examined the formation of a massive LSBG comparable
in size to Malin 1 within the IllustrisTNG simulation. The galaxy consists
of a central spheroidal component formed before $z = 0.3$ surrounded by a $> 100$ kpc
disk of gas and more recently formed stars. Its rotation curve
is also similar to what is observed for Malin 1. The authors find that the object was formed
by a merger between the galaxy's main progenitor and two other massive galaxies,
leading to stimulated accretion of gas from the progenitor's hot halo.

In this paper, we investigate the surface brightnesses of galaxies 
in the EAGLE cosmological hydrodynamical simulation, with the intent
of understanding the differences between LSBGs and HSBGs. We 
study galaxies with $M_{*} > 10^{9.5} \mathrm{M_{\odot}}$,
meaning that we do not expect our sample to include many UDGs. 
In \S\ref{methods}, we provide a brief overview of the EAGLE simulation 
and describe how we compute the
surface brightnesses of EAGLE galaxies, as well as other relevant
galaxy parameters. In \S\ref{results}, we present and discuss our results, showing
the correlations between galaxy surface brightness and other galaxy properties,
as well as the evolutionary factors that cause a galaxy to have high or low
surface brightness. Finally, in \S\ref{conclusions}, we summarize our conclusions.

Throughout this paper we assume the Planck cosmology \citep{planck}
adopted in the EAGLE simulation,
where $h = 0.6777$, $\Omega_{\Lambda} = 0.693$, $\Omega_{m} = 0.307$,
and $\Omega_{b} = 0.048$.

\section{Methods}
\label{methods}

\subsection{EAGLE simulation overview}
\label{eagle}

EAGLE \citep{eagleschaye, eaglecrain, mcalpine} is a suite of cosmological hydrodynamical simulations, run using
a modified version of the N-body smooth particle hydrodynamics (SPH) code GADGET-3 \citep{gadget}.
These modifications, described in \citet{anarchy}, are based on 
the conservative pressure-entropy formulation of SPH from \citet{hopkins}, and include 
changes to the handling of the viscosity \citep{cullen}, the conduction \citep{price}, the smoothing kernel \citep{denhen}, and 
the time-stepping \citep{durier}.

The EAGLE suite includes a number of simulation boxes 
with different sizes, resolutions, and subgrid physics prescriptions.
In this analysis we use the reference EAGLE simulation Ref-L0100N1504, 
which has box size 100 comoving Mpc per side and contains
$1504^{3}$ particles each of dark matter and baryons.
The dark matter particle mass is $9.70\times10^{6} \mathrm{M_{\odot}}$ 
and the initial gas (baryon) particle mass is $1.81\times10^{6} \mathrm{M_{\odot}}$.
The Plummer-equivalent gravitational softening length is 2.66 comoving kpc (ckpc) until $z = 2.8$ and
0.70 proper kpc (pkpc) afterward. The subgrid physics includes prescriptions for
radiative cooling, photoionization heating,
star formation, stellar mass loss, stellar feedback, 
supermassive black hole accretion and mergers, and AGN feedback.
These prescriptions and the effects of varying them are described in \citet{eagleschaye}
and \citet{eaglecrain}, and we refer the reader to these papers for details.
 
Stellar feedback in the EAGLE reference simulation was calibrated 
to approximately reproduce the local
galaxy stellar mass function (GSMF). Additionally, feedback
models that resulted in overly compact galaxy sizes at $z = 0$ 
despite reproducing the GSMF were
rejected \citep{eaglecrain}. The observed galaxy
size distributions used for comparison were derived from the SDSS \citep{shensdss} and
GAMA \citep{baldrygama} surveys. 
However, \citet{vandesande2019} found that galaxy sizes in EAGLE were too
large at fixed stellar mass compared to observations from the SAMI survey \citep{sami},
likely due to the fact that EAGLE \textit{half-mass} radii were calibrated
to observed \textit{half-light} radii. We additionally note that
neither of the two surveys to which EAGLE galaxy sizes were calibrated
extend into the low-surface brightness regime, which could also potentially
bias the resulting size distribution of EAGLE galaxies.

With only the subgrid physics calibrations described above,
EAGLE is able to reproduce a variety of properties of the observed galaxy
population. These include, among others, 
the $z = 0$ Tully-Fisher relation \citep{eagleschaye}, the evolution of the galaxy
mass function \citep{furlong2015} and galaxy sizes \citep{furlong2017}, 
optical galaxy colors and their evolution \citep{trayford2015, trayford2016},
the SFR-$M_{*}$ relation \citep{furlong2015}, and the evolution
of the star formation rate function \citep{katsianis2017}. 
Galaxy and halo catalogs as well as particle data from EAGLE have been made 
publicly available \citep{mcalpine}.

\subsection{Simulated galaxy sample}
\label{galsamp}

Galaxies in EAGLE are identified through a series of steps.
First, halos are identified in the dark matter particle distribution
using a friends-of-friends (FoF)
algorithm with a linking length of $b = 0.2$ times the mean interparticle separation \citep{davis}. 
Other particle types (gas, stars, and black holes) are assigned to 
the FoF halo of the nearest dark matter particle. 
The \textsc{subfind} \citep{subfind1, subfind2} algorithm is then
run over all the particles of any type within each FoF halo,
in order to identify local overdensities (``subhalos''). 
Each subhalo is assigned only the particles 
gravitationally bound to it, with no overlap in particles between
subhalos. The subhalo that contains the most bound particle in a FoF halo
is considered to be the ``central'' subhalo, while
any other subhalos are ``satellites''.
The stellar and gas particles bound to a subhalo are what we
consider to be an individual galaxy. Galaxy catalogs are created 
in this manner for a series of 29 simulation snapshots
from $z = 20$ to $z = 0$.

Our initial sample of simulated galaxies consists of all the galaxies 
in the Ref-L0100N1504 run of EAGLE at $z = 0$ with total bound stellar
mass $M_{*} > 10^{9.5} \mathrm{M_{\odot}}$. 
This leads to a sample of 7314 galaxies, which
includes both central and satellite galaxies. 
We compute the surface brightnesses of these galaxies using the distribution 
of their bound star particles, as described in \S\ref{galsb}.

We additionally use the galaxy merger trees from the EAGLE catalog
to investigate the evolution of galaxies with different
$z = 0$ surface brightnesses.
The merger trees were created from the simulation snapshots
using a modified version \citep{qu} of
the \textsc{d-trees} algorithm \citep{dtrees},
which assigns each subhalo a descendant in the subsequent snapshot
that contains the majority of some number of the subhalo's most bound
particles. A subhalo has only one descendant but may have multiple progenitors.
Each subhalo with at least one progenitor has a single ``main progenitor'', defined
as the one with the largest mass summed across all earlier snapshots,
as suggested by \citet{delucia} to avoid swapping of the main progenitor during major mergers.
In some cases, no descendant of a subhalo can be identified
in the next snapshot, but one can be found in a later snapshot;
because of this, descendants are identified up to 5 snapshots later.

\subsection{Galaxy surface brightness calculation}
\label{galsb}

In this subsection we describe how we obtain the surface brightness
profiles and mean surface brightnesses of the galaxies in our sample.
Physical distances are used in all calculations. The center of each
galaxy is assumed to be at the location of its most bound particle.

The aim of this paper is not to attempt a detailed statistical comparison
with LSBG observations, but rather to investigate the properties 
and evolution of the population
of objects that potentially could, depending on orientation, be observed as LSBGs
if they existed in our Universe. Thus
we compute the surface brightness for all the galaxies in our sample in the face-on
orientation, with the minor axis of the galaxy oriented along
the line of sight.

Dust-free luminosities have been computed 
for the stellar particles of EAGLE galaxies in the SDSS $ugriz$ bands 
as described in \citet{trayford2015}. Each star particle
is assumed to be a simple stellar population
 with a \citet{chabrier} initial mass function and the SPH-smoothed metallicity of
its parent gas particle. Its spectral energy distribution
is then computed using the \citet{bruzualcharlot} \textsc{galaxev} models.
Here we compute the B band luminosity of each star particle from its $u$ and $g$ band luminosities
using the transformation from Lupton 
(2005)\footnote{\href{https://www.sdss3.org/dr10/algorithms/sdssUBVRITransform.php}{www.sdss3.org/dr10/algorithms/sdssUBVRITransform.php}}:
\begin{equation}
\label{lupton}
B = u - 0.8116(u - g) + 0.1313.
\end{equation}
We compute the projected surface brightness using all the gravitationally bound
star particles within 500 kpc of the galaxy's center
(although only the most massive galaxies contain stellar particles this distant).
 
To estimate the effect of dust on the surface brightnesses of the
galaxies in our sample, we used the radiative transfer code SKIRT
\citep{skirt2003, skirt2011, skirt2015} as described
in Appendix \ref{appendix}. Overall, we find that including dust would not significantly
affect our results, and we thus neglect it for the remainder of this paper. 

To compute the surface brightness profiles, 
we first locate the 3D half-mass radius of the galaxy
using all the bound star particles. We then calculate the 
mass distribution tensor (i.e. the moment of inertia tensor) within three times
this radius. The eigenvector of this tensor corresponding to the smallest eigenvalue
is taken to be line-of-sight direction over which the luminosity is projected
to compute the surface brightness profile.

We calculate the surface brightness profile in elliptical annuli,
assuming a single axis ratio at all distances from the center
derived from the remaining two
eigenvalues/eigenvectors of the mass distribution tensor.
We compute the local surface brightness in bins of 50 particles
each. The surface brightness is converted from $\mathrm{L_{\odot}}$/pc$^{2}$ to mag/asec$^{2}$
as if all galaxies were located at $z = 0.08$, the redshift of Malin 1 \citep{impey1989}.

We would like to separate the effect of the mass-to-light ratio
on the surface brightness of galaxies from that of differences
in their stellar mass surface density. For comparison, 
we additionally compute the surface density in the form of a 
``fixed $M/L$ surface brightness profile'', using the masses of the 
stellar particles and assuming they have a fixed mass-to-light
ratio of $3 \mathrm{M_{\odot}}/\mathrm{L_{\odot}}$ in the B band. (This is
a typical $M/L_{B}$ for a passive galaxy in our sample.)

Because we consider the star particles of a galaxy to be all
those gravitationally bound to its subhalo, our surface brightness profiles
 include stellar halos and ``intracluster light'' (ICL).
This faint component is generally not considered to be part of a galaxy
for the purpose of computing the surface brightness, but it is difficult
to identify a physically-motivated boundary between the ICL and the
galaxy (see \citealt{canas2019} for a recent attempt
in cosmological hydrodynamical simulations). We discuss
the effect of this stellar component on the surface brightness
profile in \S\ref{ressb}.

\subsubsection{Mean surface brightness definition}
\label{defn}

We take a simple definition of the ``total''
surface brightness of the galaxy: the mean surface brightness within
a fixed B-band isophote of 28 mag/asec$^{2}$.
This has the advantage of not requiring fitting a parametric model
(such as a Sersic bulge and exponential disk), which we have found does not always
accurately describe the light profiles of galaxies in the simulation.
Additionally, the 28 mag/asec$^{2}$ isophote will be large for galaxies
with faint disks even if they have bright nuclei (provided that the 
disk itself is not fainter than 28 mag/asec$^{2}$), whereas this is not
necessarily the case for the half-light radius, whose position may be determined
by the central bulge. 

To obtain an estimate of the 28 mag/asec$^{2}$
isophote that avoids local fluctuations in the surface brightness, 
we smooth the surface brightness profiles computed from the particles.
This is done using a locally weighted linear regression (LOWESS; \citealt{lowess})
of the nearest 10\% of points, weighted by a tri-cube weight function:
\begin{equation}
\label{weight}
w_{i} = (1 - |d_{i}|^{3})^{3},
\end{equation}
where $|d_{i}|$ is the distance from each point in the subset
to the location at which the smoothed curve is being computed,
normalized so that the value lies between 0 and 1. 
We then take the smallest elliptical annulus at which 28 mag/asec$^{2}$ is achieved
as the isophote within which we take the average surface brightness. 
In the remainder of this paper, we will refer to the circularized radius 
of this isophote as $R_{28B}$, 
and the mean surface brightness within it as $\langle \mu_{B} \rangle$.
124 galaxies in our initial sample do not reach a surface
brightness of 28 mag/asec$^{2}$; we remove these galaxies from the sample.

We compute jackknife errors on $\langle \mu_{B} \rangle$, and
remove those galaxies for which they are larger than $0.5$ mag/asec$^{2}$.
Generally this occurs for galaxies whose surface brightness profiles
are nearly flat in the range of surface brightnesses
around 28 mag/asec$^{2}$, preventing the determination of a unique $R_{28B}$. 
This criterion further eliminates 90 of the galaxies in our sample. 

Finally, some galaxies are highly asymmetric and will be poorly
described by our assumption of elliptical isophotes. 
As noted previously, we take the center of each galaxy
to be the location of its most bound particle, which generally
corresponds to the peak of the stellar density distribution, whereas
the geometric center of the stellar distribution is
its center of mass. We thus compute
$\langle \mu_{B} \rangle$ around the galaxy center using first the star particles in the 
hemisphere directed towards the center of mass,
and then for the opposite hemisphere.
If the difference between these two values of $\langle \mu_{B} \rangle$ 
is larger than 1 mag/asec$^{2}$, we remove the galaxy from
the sample. This removes 113 galaxies, leaving us with a final sample of 6987 galaxies.

We repeat the steps above for the ``fixed $M/L$'' surface brightness profile
proportional to the surface density. 
We designate the radius of the 28 mag/asec$^{2}$ isophote identified using fixed
$M/L_{B}$ as $R_{28m}$ and the mean ``surface brightness'' within
it as $\langle \mu_{m} \rangle$.

While we discuss continuous correlations between galaxy surface brightness
and various galaxy and halo properties in this paper, 
we also define as an approximate cutoff between
LSBGs and HSBGs a value of $\langle \mu_{B} \rangle = 25.6$ mag/asec$^{2}$.
This would be the mean surface brightness within $R_{28B}$ of a purely
exponential disk having a central surface brightness of 23 mag/asec$^{2}$.

\subsection{Additional galaxy and dark matter halo properties}
\label{galproperties}

We examine the correlation between galaxy surface brightness and a 
number of properties of the galaxy and its host dark matter halo.
Many of these are precomputed values taken from
the EAGLE catalog \citep{mcalpine}, but we describe the
computation of several others here.

\subsubsection{Galaxy kinematic morphology}
\label{epsilon}

Following \citet{thob2019} and \citet{trayford2019b}, who analyzed
the kinematic morphology of EAGLE galaxies,
we compute the orbital circularity parameter \citep{abadi}
of each stellar particle:
\begin{equation}
\label{orbital_circularity}
\epsilon_{i} = j_{z,i}/j_{\mathrm{circ}}(E_{i}),
\end{equation}
where $j_{z,i}$ is the specific angular momentum
of the particle projected along the direction of total
galaxy angular momentum, and $j_{\mathrm{circ}}(E_{i})$ is the specific angular
momentum of a particle on a circular orbit with the same binding energy $E_{i}$.
The latter is estimated as the maximum
value of $j_{z,i}$ for particles with $E < E_{i}$.

We compute smoothed profiles of $\epsilon_{*}$
as a function of projected radius, using 
a weighted local linear fit of the nearest 10\% of points (similarly
as for the surface brightness profiles in \S\ref{galsb}).
This is done for our sample of galaxies at $z = 0$ as well as their
main progenitors at $z = 0.5$, in order
to examine the evolution of the kinematic distribution of the 
stellar particles as a function of radius.

We determine whether the outer regions of the galaxy are
kinematically disk-dominated using the median value of $\epsilon_{i}$ at
$R_{28B}$, denoted $\bar{\epsilon_{*}}(R_{28B})$. 
We find that this value generally correlates
well with the median value of $\epsilon_{i}$ 
for the entire galaxy. However, some galaxies
with large central bulges but very faint, extended disks
can have low overall $\bar{\epsilon_{*}}$ but high values at $R_{28B}$.

For most galaxies, $\epsilon_{*}(R_{28B}) = 0.5$ corresponds approximately to
$\bar{\epsilon_{*}} = 0.3$ for the entire galaxy, which is the threshold between
disk- and bulge-dominated galaxies recommended by \citet{thob2019}
based on the division between passive and star-forming galaxies 
found in \citet{correa2017}. We thus use $\epsilon_{*}(R_{28B}) > 0.5$
as a threshold for kinematic disk dominance at $R_{28B}$.

\subsubsection{Ex-situ stellar mass fraction}
\label{fexsitu}

To quantify the impact of mergers on the galaxies in our sample,
we estimate the fraction of each galaxy's stellar mass formed outside
of the galaxy --- the ex-situ stellar mass fraction. 

Galaxies merging into a more massive galaxy are often stripped of their outer stars prior
to the simulation snapshot at which the merger event is recorded. These stripped 
stars join the more massive galaxy prior to the merger, and thus are not 
recorded as part of the mass merging into the galaxy during the merger event. As a result,
the mass of a merging satellite at the time of a merger 
can be a significant underestimate of the contribution of ex-situ stars.

We instead estimate the ex-situ mass fraction as follows. For each stellar particle in EAGLE,
the time at which it formed from its parent gas particle is recorded. These timesteps
have much finer spacing than the spacing of the snapshots
used in our galaxy merger trees. For each stellar particle within a galaxy at $z = 0$,
if in the snapshot immediately after its formation it is bound to the main progenitor
of the $z = 0$ galaxy, we consider it to be part of the in-situ stellar mass. 
Otherwise, it contributes to the ex-situ mass fraction. 

This method still somewhat underestimates the fraction of ex-situ stellar mass,
because stars that form in a galaxy less than one snapshot ($\le 1.35$ Gyr) before
it merges into a more massive galaxy will be counted as in-situ mass. 
However, we find that this estimate of ex-situ mass is still larger than using 
the masses of non-main progenitors at the time of a merger.

\subsubsection{Matched dark matter only halos}
\label{dmo}

We would like to investigate whether the properties
of dark matter halos influence the surface brightness
of the galaxies that form within them. However, galaxies are also able to alter the properties
of their host dark matter halos (e.g., \citealt{schaller2015}), leading
to difficulty separating cause from effect when examining correlations
between galaxy and halo properties. For this reason, we use halos from
the matching dark matter only (DMO) run of the EAGLE simulation.

The DMO simulation has identical box size, resolution, and initial conditions
as the reference EAGLE run. It contains (1504)$^{3}$ particles of dark matter, 
each with mass $1.15\times10^{7} \mathrm{M_{\odot}}$.
Each particle in the reference and DMO runs is tagged with a unique ID based
on its initial conditions, such that the equivalent particles can
be identified in both simulations. To find corresponding dark matter subhalos between
the two simulations, we use the method described in \citet{schaller2015}, which 
considers two subhalos to be ``equivalent''
if over half of the 50 most bound particles of each one are also bound to
the other.

Because dark matter halos that become satellites are subject to stripping, which substantially 
alters their properties, and because some halo properties (e.g. $M_{200}$; see list below) are ill-defined
for satellite subhalos, we examine only the properties of central 
galaxies/subhalos relative to their surrounding FoF halo. 
Of the galaxies in our sample, 4098 are hosted by central subhalos in the reference simulation, 
and 3826 (93.4\%) of these are successfully matched to central subhalos in the DMO EAGLE run.

The dark matter halo properties we examine include:
\begin{itemize}
\item $M_{200c}$, the FoF halo mass within 
 $r_{200c}$, the radius at which the mean internal density is equal to 200 times the critical
density of the Universe.
\item $V_{\mathrm{max}}/V_{200c}$, where $V_{\mathrm{max}}$ is the 
central subhalo maximum circular velocity, and $V_{200c} = \sqrt{GM_{200c}/r_{200c}}$.
This quantity serves as a proxy for the halo concentration \citep{prada2012}.
\item the halo spin parameter of \citet{bullock2001}:
\begin{equation}
\label{bullock}
\lambda = \frac{J}{\sqrt{2}MrV},
\end{equation}
where $J$ is the total angular momentum and $M$ is the total mass 
within some radius $r$, and $V = \sqrt{GM/r}$. We compute
the spin parameter within $r_{200c}$ as well as $0.1r_{200c}$, in order to represent
the spin of the entire and the ``inner'' halo, respectively.
\end{itemize}

\begin{figure}
  \begin{center}
    \includegraphics[width=\columnwidth]{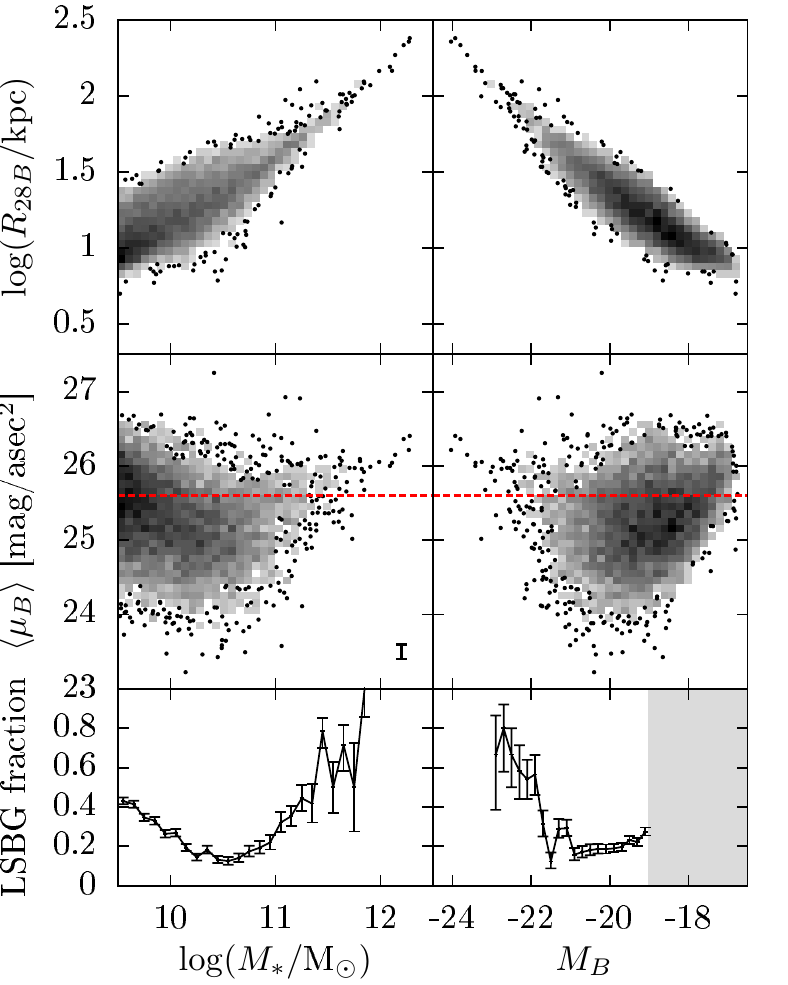}
    \caption{The distribution of radii and surface brightnesses for our sample of galaxies.
The shading of each bin is proportional to the log-number of galaxies in the bin. Bins with 
fewer than 3 galaxies are plotted with a separate point for each galaxy.
\textit{Top left:} The circularized 28 mag/asec$^{2}$ isophotal radius in the B band, $R_{28B}$,
as a function of galaxy stellar mass. \textit{Top right:} $R_{28B}$ as a function of the galaxy 
magnitude in the B band, $M_{B}$. \textit{Center left:} The mean surface brightness within $R_{28B}$,
$\langle \mu_{B} \rangle$, versus galaxy stellar mass. The error bar on the bottom right shows the
median size of the error on $\langle \mu_{B} \rangle$, which is $\pm 0.099$ 
mag/asec$^{2}$. (The median error on $\log(R_{28B}/\mathrm{kpc})$ is $\pm 0.012$, which is smaller than the size of a plotted bin.)
The red dashed line denotes our adopted division between LSBGs and HSBGs (see \S\ref{defn}). 
\textit{Center right:} $\langle \mu_{B} \rangle$ as a function of $M_{B}$.
\textit{Bottom left:} The fraction of galaxies that we classify as low surface brightness as a function of $M_{*}$. The fraction of LSBGs is high at low masses and decreases significantly at higher masses. An upturn in the LSBG fraction is visible for $M_{*} > 10^{11} \mathrm{M_{\odot}}$, which is due to the
buildup of stellar halos with $\langle \mu_{B} \rangle < 28$ mag/asec$^{2}$ around high-mass galaxies;
see \S\ref{ressb} for discussion.
\textit{Bottom right:} The fraction of LSBGs as a function of $M_{B}$. Because our sample is selected by a cut in stellar mass, it is incomplete for $M_{B} > -19$; this region is shaded gray.
}
    \label{fig1}
  \end{center}
\end{figure}

\begin{figure}
  \begin{center}
    \includegraphics[width=\columnwidth]{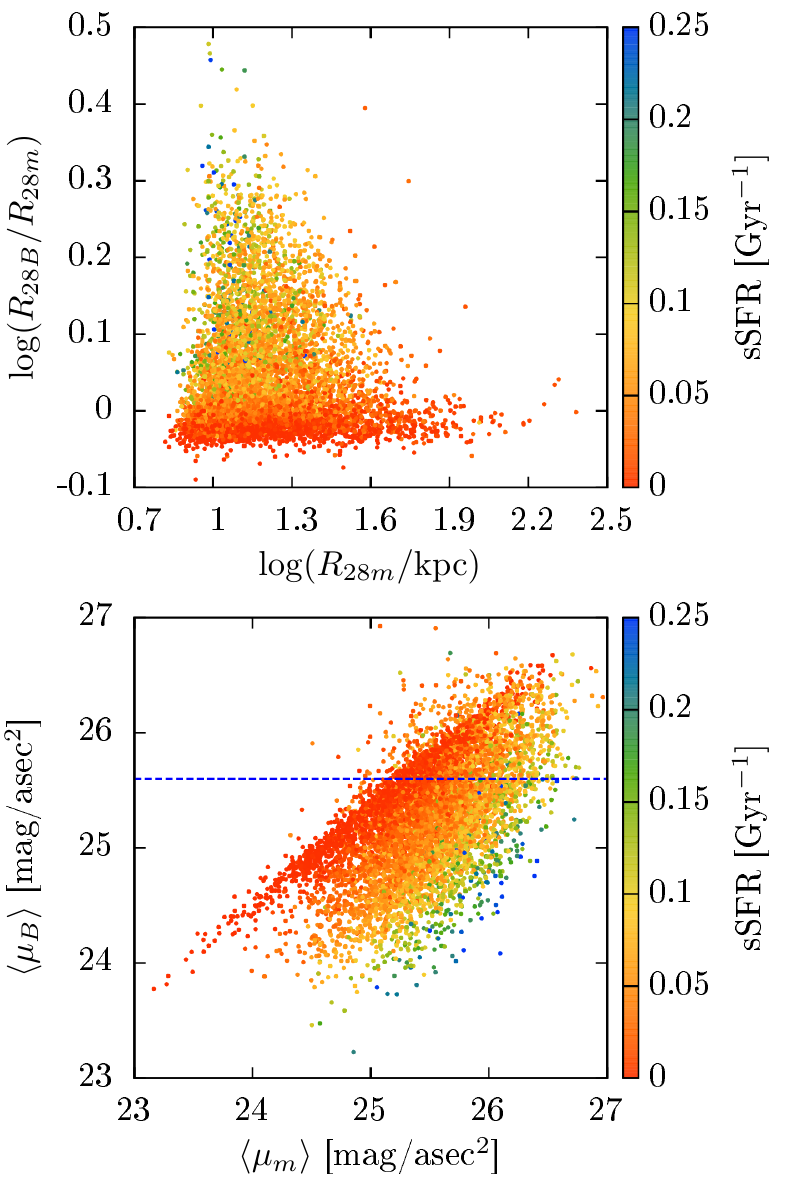}
    \caption{\textit{Top panel:}
The horizontal axis shows $R_{28m}$, the circularized radius of the 28 mag/asec$^2$ B-band isophote that
each galaxy would have if $M/L_{B} = 3 \mathrm{M_{\odot}}/\mathrm{L_{\odot}}$. On the vertical axis is the ratio of
$R_{28B}$ to $R_{28m}$; the true value of $M/L_{B}$ for most galaxies is such that $R_{28B}/R_{28m} \gtrsim 1$.
The color coding corresponds to the specific star formation rate (sSFR) of each galaxy, showing that galaxies
with higher sSFR have larger $R_{28B}$ at fixed $R_{28m}$, as expected from their
lower mass-to-light ratios and consequently larger luminosities.
\textit{Bottom panel:}
$\langle \mu_{m} \rangle$, the mean surface brightness if all galaxies had 
$M/L_{B} = 3 \mathrm{M_{\odot}}/\mathrm{L_{\odot}}$, versus
the true mean surface brightness, $\langle \mu_{B} \rangle$. There is a correlation between surface brightness
and surface mass density, but with a substantial scatter. 
The colors again represent sSFR. Although a higher sSFR increases $R_{28B}$, 
there is a strong trend for galaxies
with high sSFR to also be brighter within this radius at fixed mass surface density. The dashed blue line demarcates
$\langle \mu_{B} \rangle = 25.6$ mag/asec$^{2}$, as in Figure \ref{fig1}. It can be seen that LSBGs 
selected in the B band comprise galaxies with very low mean surface density as well as those with typical
surface densities but very low star formation rates.
}
    \label{fig3}
  \end{center}
\end{figure}

\begin{figure*}
  \begin{center}
    \includegraphics[width=\textwidth]{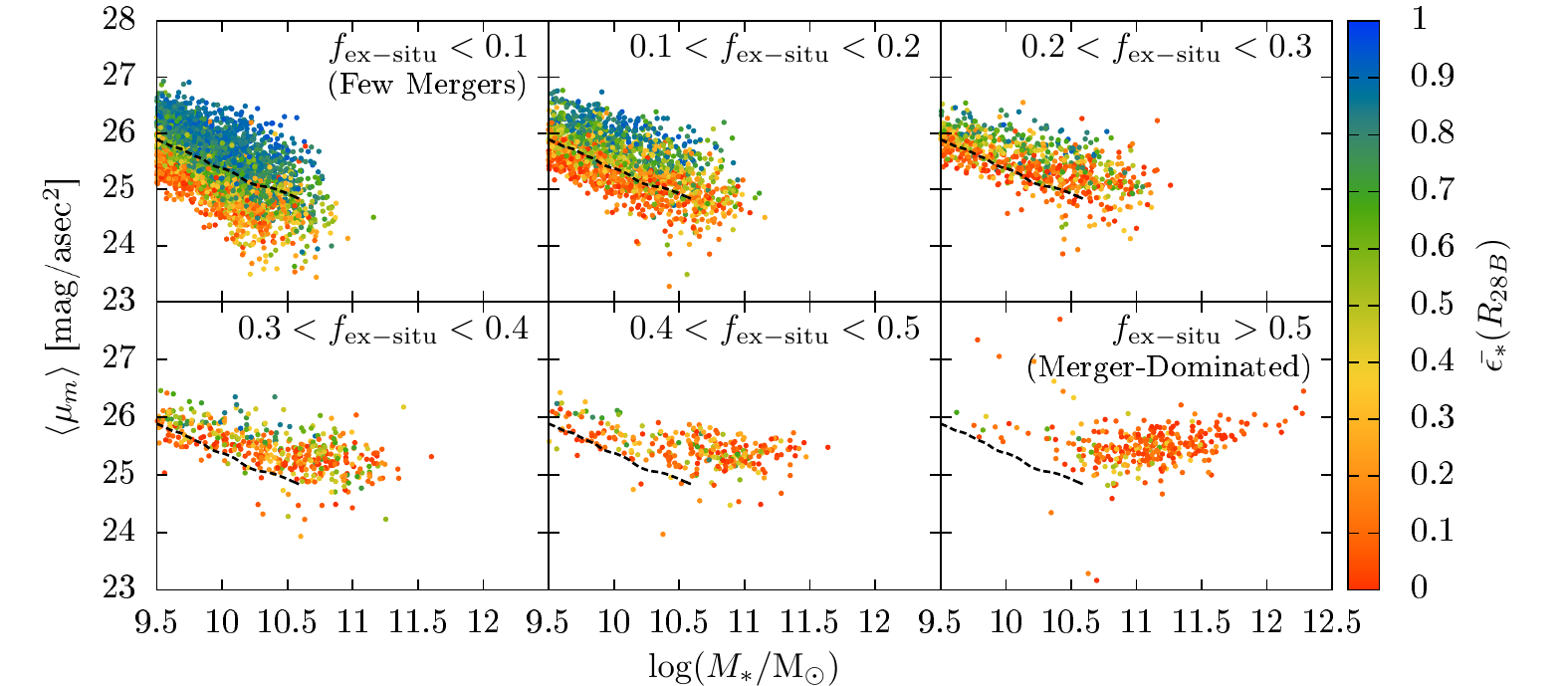}
    \caption{The relation between galaxy stellar mass and mean stellar mass surface density
(expressed as $\langle \mu_{m} \rangle$) for bins of different ex-situ stellar mass fraction ($f_{\mathrm{ex-situ}}$):
the fraction of each galaxy's stellar mass formed outside of its main progenitor branch. $f_{\mathrm{ex-situ}}$ is higher
for galaxies that have gained more of their stellar mass from mergers. 
The dashed black line shows the median $M_{*}-\langle \mu_{m} \rangle$ relation for $f_{\mathrm{ex-situ}} < 0.1$,
and is repeated in each panel to facilitate comparison. Mergers tighten
the relation between $M_{*}$ and $\langle \mu_{m} \rangle$ and lower the median surface density
at fixed $M_{*}$. The latter effect is more pronounced towards higher $M_{*}$.
This is a result of the buildup of the diffuse stellar halo/intracluster light component
surrounding the galaxy, which lowers the \textit{mean} surface density, as explained in \S\ref{ressb}.
The color coding shows the median orbital circularity
parameter at $R_{28B}$, a measure of how kinematically
rotation-dominated the galaxy is at $R_{28B}$ (\S\ref{epsilon}). For galaxies with few
mergers ($f_{\mathrm{ex-situ}} < 0.1$), $\langle \mu_{m} \rangle$ is essentially a function
of $M_{*}$ and $\bar{\epsilon_{*}}(R_{28B})$. For galaxies containing an increasing fraction of 
stellar mass from mergers, the kinematic morphologies become more uniformly dispersion-dominated ($\bar{\epsilon_{*}}(R_{28B}) \approx 0$).
}
    \label{fig4}
  \end{center}
\end{figure*}

\begin{figure*}
  \begin{center}
    \includegraphics[width=\textwidth]{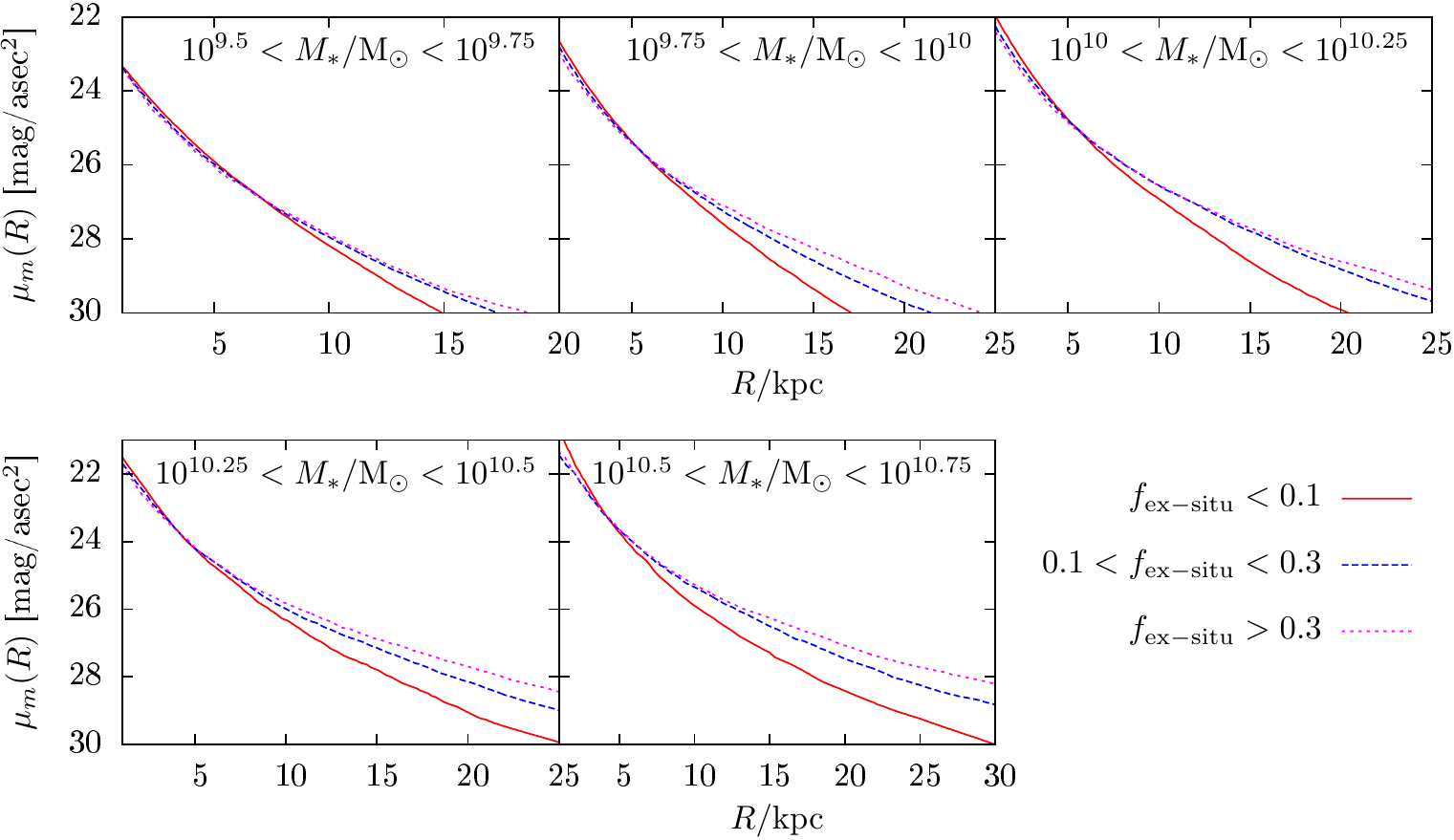}
    \caption{Median $\mu_{m}$ (stellar mass surface density divided by $3 \mathrm{M_{\odot}}/\mathrm{L_{\odot}}$) 
profiles in projected radius $R$. Different lines in each panel are profiles
for galaxies that have gained different fractions of their $z = 0$ stellar mass from mergers:
red solid lines represent $f_{\mathrm{ex-situ}} < 0.1$, blue dashed lines $0.1 < f_{\mathrm{ex-situ}} < 0.3$,
and magenta dotted lines $f_{\mathrm{ex-situ}} > 0.3$.
Panels show bins of galaxy stellar mass. 
To ensure that the different surface brightness profiles of galaxies
with differing $f_{\mathrm{ex-situ}}$ do not merely result
from the fact that galaxies that have experienced more mergers are more likely
to be spheroidal, we subsample the
galaxies in bins of $f_{\mathrm{ex-situ}}$
such that they have the same distribution of $\bar{\epsilon_{*}}$, the median
 orbital circularity parameter computed for all the star particles in the galaxy. 
It can be seen that mergers
cause a diffuse stellar halo to develop in the outer parts of galaxies in all mass bins.
The surface density at which the profiles diverge is larger at higher masses.}
    \label{fig5}
  \end{center}
\end{figure*}

\begin{figure*}
  \begin{center}
    \includegraphics[width=\textwidth]{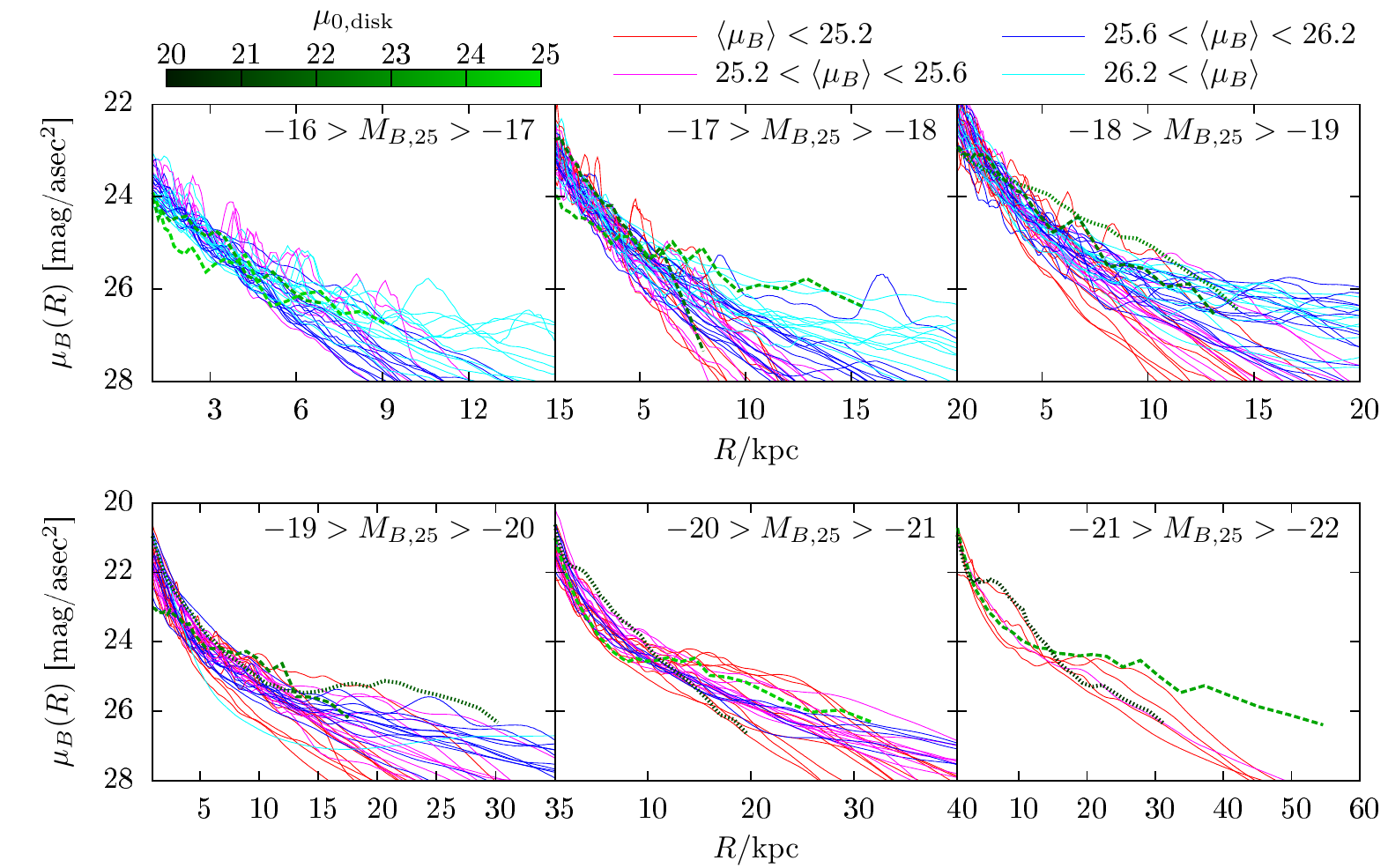}
    \caption{Radial surface brightness profiles of simulated galaxies in our sample, compared
    to observed profiles of low surface brightness profiles taken from the literature. 
The different panels represent bins in galaxy magnitude, measured within the 25 mag/asec$^{2}$ isophote, so as to be more easily compared to observations.
 In each panel, smoothed surface brightness profiles of individual simulated 
galaxies are shown, 
selected to be equally sampled in magnitude and $\langle \mu_{B} \rangle$.
To avoid comparison to the surface brightness profiles of spheroidal halos,
the galaxies shown are limited to those with $\bar{\epsilon_{*}}(R_{28B}) > 0.5$.
 Red, magenta, blue, and cyan lines represent galaxies of increasingly faint $\langle \mu_{B} \rangle$, as indicated in the legend. The observed LSBG profiles are
represented by green lines of varying shades. They are taken from \citet{galazlsbg}
and \citet{mcgaughlsbg}, with the former represented by dotted lines and the latter
by dashed lines. The shade of the line represents the central surface brightness
of the galactic disk. Each panel shows the LSBGs with the brightest and faintest
disk central surface brightness from the two papers mentioned above, so as to bracket
the diversity of LSBG profiles. The literature 
profiles have been corrected to the Planck cosmology
used in this paper as well as adjusted to the same redshift ($z = 0.08$) assumed
for our simulated profiles.
}
    \label{fig6}
  \end{center}
\end{figure*}

\begin{figure}
  \begin{center}
    \includegraphics[width=\columnwidth]{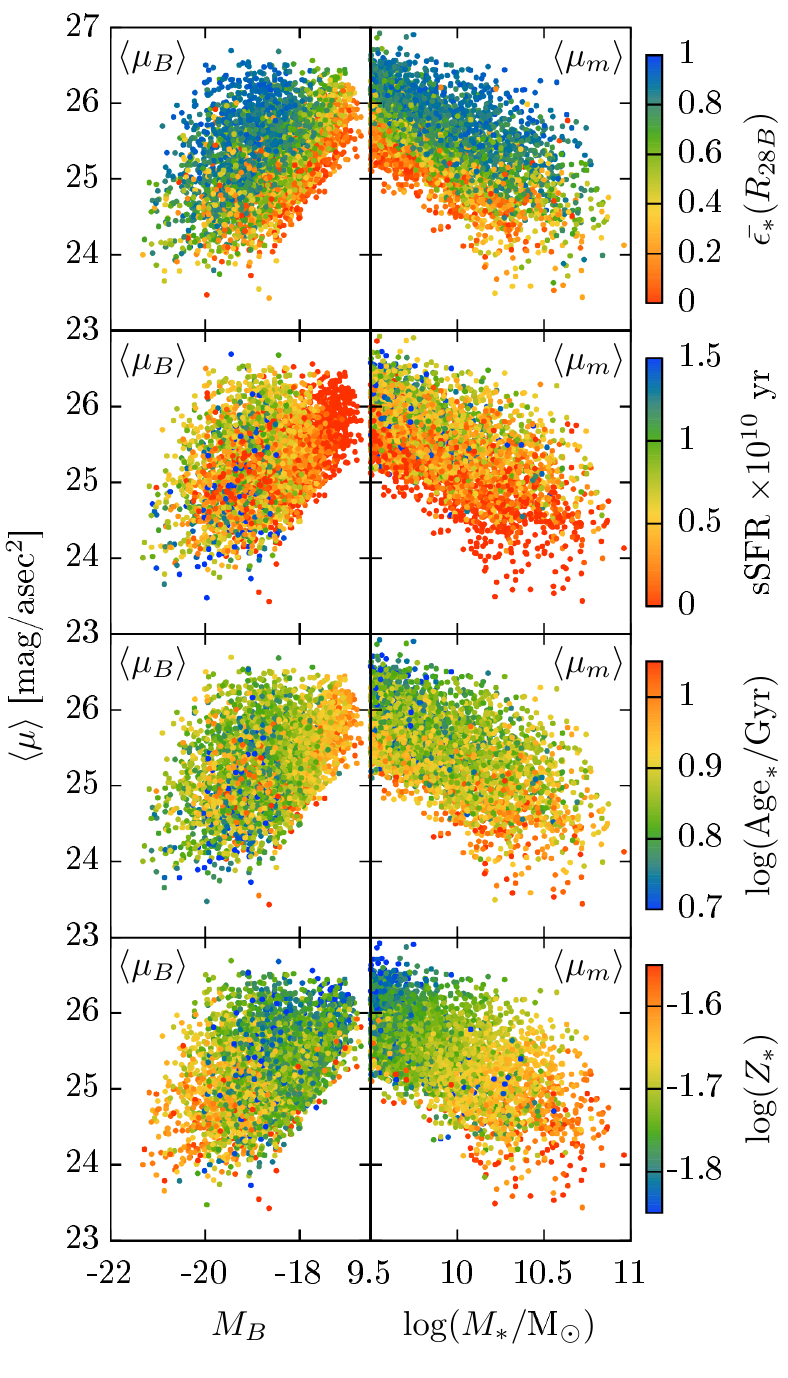}
    \caption{For galaxies with $f_{\mathrm{ex-situ}} < 0.1$, 
the correlation of various galaxy properties with $M_{B}$ and $\langle \mu_{B} \rangle$ in
the left column, and with $M_{*}$ and $\langle \mu_{m} \rangle$ in the right column. 
The color coding is as follows.
\textit{Top row:} The median orbital circularity parameter, 
$\bar{\epsilon_{*}}$, at $R_{28B}$, a measure of the kinematic morphology of 
the external regions of each galaxy.
The right-hand panel is the same as the top left-hand panel of Figure \ref{fig4}. The tight relation
between kinematic morphology and stellar mass surface density at fixed stellar mass produces a similarly
tight relation between $M_{B}$, $\langle \mu_{B} \rangle$, and $\bar{\epsilon_{*}}(R_{28B})$.
\textit{Second row:} Specific star formation rate (sSFR). Lower
surface density correlates with higher sSFR (right panel). However, surface brightness
in the B band is positively correlated with a high sSFR (see Figure \ref{fig3}), 
so there is little correlation between sSFR and $\langle \mu_{B} \rangle$ (left panel).
The increase in $\langle \mu_{B} \rangle$ at low luminosity is a selection
effect due to the fact that we select galaxies with a fixed $M_{*}$ cut, so
the least luminous galaxies are those with low sSFR (high $M/L_{B}$).
\textit{Third row:} The initial-mass-weighted mean stellar age of each galaxy. Trends are 
similar to those with sSFR.
\textit{Bottom row:} The stellar metallicity (metal mass fraction) of each galaxy, $Z_{*}$. There is
 a tight relation between $\langle \mu_{m} \rangle$ and stellar metallicity at fixed $M_{*}$
seen in the right panel. This correlation persists, albeit with more scatter, between
$Z_{*}$ and $\langle \mu_{B} \rangle$ at fixed $M_{B}$, such that lower surface brightness
galaxies are more metal-poor.}
    \label{fig7}
  \end{center}
\end{figure}

\section{Results}
\label{results}

\subsection{Distribution of galaxy surface brightnesses}
\label{ressb}

In Figure \ref{fig1}, we present the distribution of 28 mag/asec$^{2}$
isophotal radii and mean surface brightnesses for our $z = 0$ galaxy sample.
The top row presents $R_{28B}$, the circularized (projected) radius of the elliptical isophote
corresponding to 28 mag/asec$^{2}$ in the B band. The left column
shows this value as a function of total galaxy stellar mass ($M_{*}$),
and the right column as a function of the total galaxy B-band magnitude ($M_{B}$).
Note that both the stellar mass and galaxy luminosity in Figure \ref{fig1}
are the \textit{total} mass/luminosity
of \textit{all} the star particles bound to the galaxy's subhalo. This is in contrast
to most other papers using the EAGLE simulation, which measure these 
quantities within a 3D aperture of 30 pkpc (e.g. \citealt{eagleschaye}), but in doing so exclude
any contribution from extended parts of the galaxy.

We see that $R_{28B}$ has a strong positive correlation with galaxy B-band luminosity,
whereas the relationship between $R_{28B}$ 
and $M_{*}$ is less tight, particularly at lower masses.
This is due to the scatter in galaxy mass-to-light ratio in the B band
at fixed stellar mass.

In the middle panels, we show the correlation of $\langle \mu_{B} \rangle$
with $M_{*}$ and $M_{B}$. We place a dashed red line
at 25.6 mag/asec$^{2}$, which we use as a division
between LSBGs and HSBGs (see \ref{defn}).
$R_{28B}$ typically encloses $\approx 95\%$ of the total galaxy light, and as a result,
$\langle \mu_{B} \rangle$ is largely determined by the
size of $R_{28B}$ at fixed $M_{*}$ or $M_{B}$.
Low surface brightness galaxies are those that
are more extended in their light profiles at fixed luminosity.

The bottom panels of Figure \ref{fig1} show the fraction of galaxies that 
we classify as LSBGs as a function of $M_{*}$ and $M_{B}$.
Because our sample is selected based on a stellar mass cut, it is incomplete
for $M_{B} > -19$. It can be seen that there 
are significantly more LSBGs at low masses and faint magnitudes. This
is not surprising given that most low surface brightness galaxies
in observations are faint, low-mass galaxies (e.g. \citealt{dalcanton}).

At $M_{B} = -19$, $\approx 20\%$ of the galaxies in our sample are classified
as low surface brightness. This is higher than the observed fraction
reported by \citet{lumfunc} at this galaxy magnitude, although is within
their substantial uncertainty. However, our definition of LSBGs is not identical to
theirs, as they identify LSBGs by central surface brightness, whereas
we classify those with substantial bulges but very faint extended disks as LSBGs as well;
therefore, it is not surprising that our galaxy sample is somewhat
different from theirs.

There is a noticeable upturn towards fainter $\langle \mu_{B} \rangle$
at the highest masses ($\gtrsim 10^{11} \mathrm{M_{\odot}}$). 
This results from the fact that the stars considered to belong to a galaxy
in EAGLE are all those which are gravitationally bound to it. This includes
their stellar halos and, for the most massive galaxies,
a significant fraction of the diffuse ``intracluster light'' component.
We will discuss this in detail later in this subsection.

The surface brightness of galaxies is, naturally,
influenced by their mass-to-light ratio.
In the top panel of Figure \ref{fig3}, we compare $R_{28B}$ to $R_{28m}$ computed for
galaxies using a fixed mass-to-light ratio of $3 \mathrm{M_{\odot}}/\mathrm{L_{\odot}}$.
The color coding indicates the specific star formation rate (sSFR)
of each galaxy. There is a clear trend such that galaxies with more star
formation have larger 28 mag/asec$^{2}$ isophotes than would be expected
from their stellar mass surface density profiles with the approximate $M/L$
of a passive galaxy. This is due to the fact that sSFR is tightly correlated
with the mass-to-light ratio in the B band, a blue band that is sensitive to
the presence of young stars.

We further see the influence of the sSFR in the bottom panel
of Figure \ref{fig3},
which shows $\langle \mu_{B} \rangle$ versus $\langle \mu_{m} \rangle$.
At fixed stellar mass surface density, a higher
sSFR correlates strongly with a brighter galaxy, despite
the fact that it also correlates with a larger $R_{28B}$.
It is clear from the lower panel that, for the majority of galaxies, 
the B-band surface brightness is
essentially determined by a combination of galaxy surface density and 
sSFR. A dashed line is again drawn at 25.6 mag/asec$^{2}$; galaxies
lying above this line include both those with very low surface densities
and those with more typical surface densities but very low sSFR.
In EAGLE, the star formation rates of galaxies have been
found to vary on both long and short timescales \citep{matthee2019}.
Given the correlation 
between sSFR and surface brightness seen in Figure \ref{fig3}, the surface brightnesses
of galaxies likely also exhibit some short-timescale 
variation.
 
Having seen the correlation between $\langle \mu_{B} \rangle$ and $\langle \mu_{m} \rangle$, we now
turn back to the influence of stellar halos/ICL. In Figure \ref{fig4}, we plot
the galaxy stellar mass versus $\langle \mu_{m} \rangle$, in bins of different
ex-situ stellar mass fraction. Galaxies with $f_{\mathrm{ex-situ}} < 0.1$ have
had little influence from mergers. From left to right and top to bottom,
galaxies with an increasing stellar mass contribution from mergers are shown. The color coding shows 
the median orbital circularity parameter at $R_{28B}$, 
$\bar{\epsilon_{*}}(R_{28B})$, a measure of how kinematically
rotation-dominated the galaxy is at that radius
(see \S\ref{epsilon}). For galaxies
with little contribution from mergers, the stellar mass surface density at fixed $M_{*}$
is largely a function of the kinematic morphology, with galaxies having
large disks being less dense. As $f_{\mathrm{ex-situ}}$ increases,
a larger fraction of galaxies are dispersion-dominated, and the
correlation between $\langle \mu_{m} \rangle$ and kinematic morphology gradually disappears.
At fixed stellar mass, the distribution of $\langle \mu_{m} \rangle$ becomes tighter,
and its mean value becomes larger (lower density). The latter effect is increasingly
pronounced for larger stellar masses.

Figure \ref{fig5} reveals the reason for some of these trends with ex-situ
mass fraction. Here we plot the median $\mu_{m}(R)$ profiles
for galaxies in different stellar mass bins and with different ex-situ stellar mass fractions.
In each mass bin, we subsample the galaxies with different $f_{\mathrm{ex-situ}}$
such that they have the same distribution in $\bar{\epsilon_{*}}$, 
the median orbital circulary parameter of all the star particles in the galaxy; thus the
differences in their stellar mass density profiles are not simply the result
of different mean kinematic morphology.
At small radii, the $\mu_{m}$ profiles of
galaxies with different ex-situ mass fractions are similar.
However, at all masses, galaxies that have experienced more mergers have 
more extended low-density (faint) outer parts.
This faint component corresponds to the stellar halo, or, for the most massive
galaxies, the ``intracluster light''. Because the \textit{fraction}
of mass contained in this component is largely determined by $f_{\mathrm{ex-situ}}$,
the stellar halo dominates the mass surface density profile
beginning at higher surface densities with increasing stellar mass,
as can be seen by comparing the panels of Figure \ref{fig5}. As a result, 
the value of $R_{28m}$ is increased more for higher-mass galaxies, and the
mean surface density within $R_{28m}$ is correspondingly decreased.
This is responsible for the trend in Figure \ref{fig4} whereby
mergers increase the mean value of $\langle \mu_{m} \rangle$ more
for galaxies with higher stellar masses. It is similarly 
responsible for the upturn in $\langle \mu_{B} \rangle$
seen at high masses in Figure \ref{fig1}. We note that the magnitude
of this effect is sensitive to the particular choice of isophote
within which the mean surface brightness is measured,
as this determines the galaxy mass at which the stellar halo begins to 
influence the location of the isophote.

Finally, we compare the surface brightness profiles
of our simulated galaxies to those of observed low surface brightness galaxies
in Figure \ref{fig6}. Each panel represents a bin in galaxy magnitude
within the 25 mag/asec$^2$ isophote; this limit was chosen to allow for better
comparison to observed surface brightness profiles of LSBGs, which generally
do not extend significantly fainter than this limit. 
In each panel, we plot the surface brightness
 profiles of a selection of simulated galaxies chosen to span the range in galaxy magnitude
and $\langle \mu_{B} \rangle$ in that magnitude bin. 
To avoid comparing observed LSBGs to the surface brightness profiles
of faint halos that were seen in Figure \ref{fig5}, we show
only those galaxies that are disk-dominated in their
outer regions ($\bar{\epsilon_{*}}(R_{28B}) > 0.5$).
The color-coding represents the value of $\langle \mu_{B} \rangle$, 
with profiles colored blue and cyan representing 
galaxies that we classify as low surface brightness.
There is a substantial diversity of profile shapes in each panel, driven
both by differences in surface brightness as well as the range in galaxy
magnitude shown in each panel. We note that our galaxy sample is incomplete
in the top row of panels, because we select the sample based on a mass cut. This 
excludes bright galaxies with low mass. Thus the top row of panels
does not show the full range of galaxy surface brightnesses that are actually present
at faint magnitudes.

For comparison to observations, in each panel we plot observed low surface brightness
galaxies from \citet{galazlsbg} and \citet{mcgaughlsbg}. We show only those galaxies 
that were successfully fit in these papers with an exponential disk component (possibly
in addition to a bulge), 
and we represent the central surface brightness of this component ($\mu_{0}$) by plotting
the profiles in different shades of green. In each panel, 
we plot the observed LSBGs in that magnitude bin with the 
brightest and faintest central surface brightnesses, so as to bracket the observations.
We see that the disk central surface brightness does not appear 
to correspond to a fixed $\langle \mu_{B} \rangle$ as a function of magnitude; this 
is likely due to the influence of bulges at higher galaxy masses on $\langle \mu_{B} \rangle$.
Within each magnitude bin, the 
observed LSBGs do tend to look more similar to those simulated galaxies 
with fainter $\langle \mu_{B} \rangle$. Given that
the profiles of the observed LSBGs generally extend to $\sim$25 mag/asec$^{2}$, 
whereas our galaxy sample is chosen based on the 28 mag/asec$^{2}$, it is not
entirely surprising that the samples differ somewhat. Nevertheless,
we find that the diversity of surface 
brightness profiles present in the EAGLE simulation generally
encompasses the shapes of observed LSBG surface brightness profiles.

\begin{figure}
\begin{center}
  \includegraphics[width=\columnwidth]{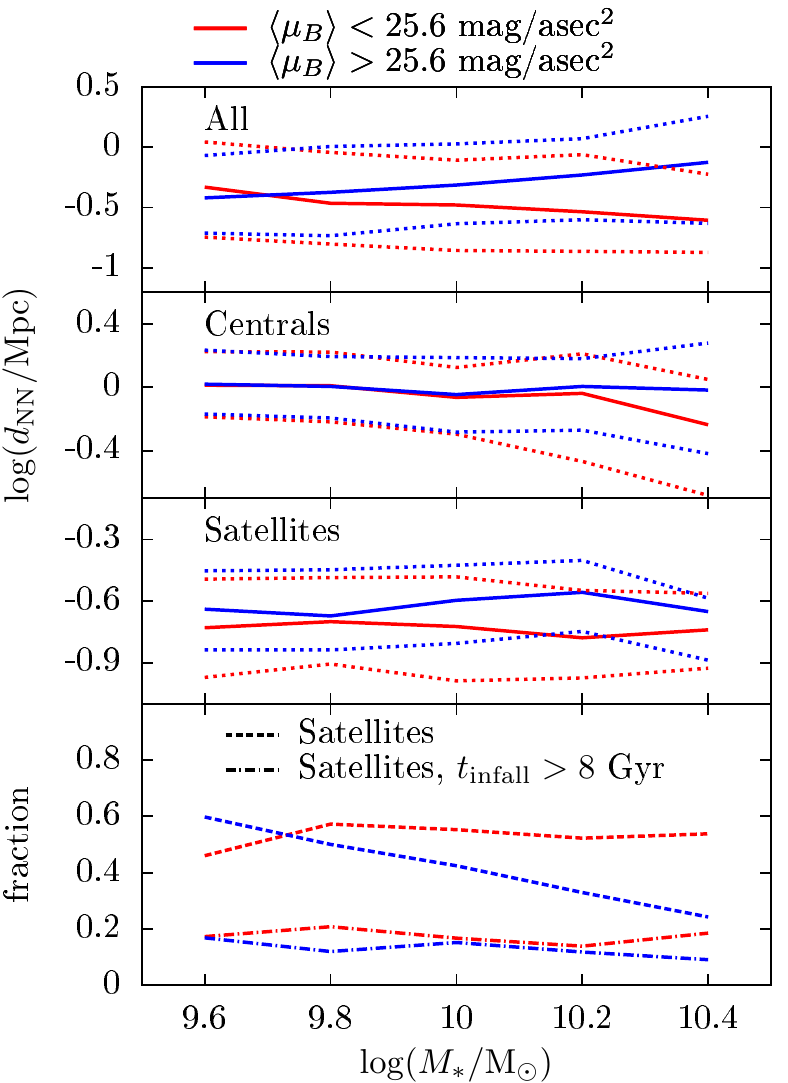}
  \caption{Properties related to the spatial distribution of galaxies
as a function of stellar mass, for galaxies with $f_{\mathrm{ex-situ}} < 0.1$,
split into LSBGs (blue) and HSBGs (red).
\textit{Top panel:} Distance from the nearest neighbor galaxy, $d_{\mathrm{NN}}$,
where a ``neighbor'' is any galaxy with $M_{*} > 10^{9} \mathrm{M}_{\odot}$. 
Solid lines represent the median value and dotted lines the top and bottom quartiles.
It is apparent that LSBGs tend to be farther away from their nearest neighbor
than HSBGs at all but the lowest stellar masses.
\textit{Second panel:} Same as the top panel, but only for central galaxies. Note
that some centrals have no satellites above our mass cut --- their nearest neighbor
is in a different FoF halo. 
In contrast to the top panel, LSBG and HSBG central galaxies the same
distribution of $d_{NN}$, except at the highest masses.
\textit{Third panel:} Same as top two panels, but only for satellite galaxies.
LSBG satellites tend to be farther from their nearest neighbor than HSBG satellites
at all stellar masses.
This is especially apparent in the lower quartile of their $d_{NN}$ distributions,
suggesting that LSBGs are particularly unlikely to have very nearby neighboring
galaxies.
\textit{Bottom panel:} Dashed lines represent the fraction of LSBGs and HSBGs
that are satellite galaxies in each stellar mass bin.
HSBGs are more likely to be satellites at all but the lowest masses. Dot-dashed lines
show the fraction of galaxies in each bin that are satellites that fell
into their host halo after a cosmic age of 8 Gyr (i.e. that became
satellites comparatively recently). Here we see that LSBGs and HSBGs
are about equally likely to be ``young'' satellites. This implies that 
the difference in overall satellite fraction is due to
the fact that HSBGs are more likely to be satellites that fell into their
hosts long ago, and have had more time to be disrupted.
}
\label{fig8}
\end{center}
\end{figure}

\subsection{Surface brightnesses of galaxies with low ex-situ mass fractions}

\label{resnomerge}

In Figure \ref{fig4} and Figure \ref{fig5}, galaxy mergers
were shown to have two typical effects: they cause galaxies to become more kinematically spheroid-dominated,
and they build up a diffuse stellar halo in the outskirts of the galaxy.
Nevertheless, a minority of galaxies with extended kinematic disks 
($\bar{\epsilon_{*}}(R_{28B} > 0.5$)
are visible even in the panels with high ex-situ mass fractions in Figure \ref{fig4}.
We will demonstrate in \S\ref{resmergers} that in a minority of cases, mergers
can in fact increase the spin of galaxies
 as well as contribute to a faint outer \textit{disk} rather than a spheroidal
halo. However, in this section we will focus on the formation of LSBGs with
low ex-situ mass fractions, as it is clear from Figure \ref{fig1} that
the majority of LSBGs are low-mass galaxies, which tend to
have relatively quiescent merger histories (Figure \ref{fig4}).

We focus specifically on galaxies with an ex-situ
stellar mass fraction less than 0.1. This subsample contains galaxies with stellar masses $M_{*} \lesssim
10^{10.75} \mathrm{M_{\odot}}$, and comprises $\approx 60\%$ of galaxies with these masses.
While mergers clearly alter the properties of the galaxy population, the
magnitude of the effect depends on how much stellar mass they contribute,
as can be seen from the sequence of panels in Figure \ref{fig4}. Although the following
three subsections focus on a subsample of galaxies with very little influence from mergers, 
our qualitative conclusions would be unchanged
if we instead considered the all galaxies with $f_{\mathrm{ex-situ}} < 0.2$,
which includes $81\%$ of all galaxies with $M_{*} < 10^{10.75} \mathrm{M_{\odot}}$, although
the scatter in the correlations we identify would be larger.
We therefore note that the LSBG properties and formation scenario identified in the 
following subsections likely apply to the considerable majority of low-mass LSBGs,
which also dominate the overall LSBG population for $M_{*} > 10^{9.5} \mathrm{M_{\odot}}$.

In Figure \ref{fig7}, we show in the left column of panels the galaxy B-band magnitude
$M_{B}$ versus $\langle \mu_{B} \rangle$. In the right column we show the galaxy stellar mass
$M_{*}$ versus $\langle \mu_{m} \rangle$. In the top row,
the colors show the median orbital circularity parameter at $R_{28B}$, $\bar{\epsilon_{*}}(R_{28B}$, as in Figure \ref{fig4}. The correlation between the orbital circularity and surface density
also extends to surface brightness, such that low surface brightness galaxies are 
more rotation dominated. This agrees with observations of LSBGs, 
which typically have disky morphologies.

In the second row of panels in Figure \ref{fig7}, the color coding represents the specific
star formation rate. In the right panel, we see that galaxies with lower surface densities
have higher sSFR at fixed stellar mass. This is unsurprising given the tight correlation
between surface density and kinematic morphology, as rotation-dominated galaxies in EAGLE
are known to have higher star formation rates \citep{correa2017}. However,
since a higher sSFR also increases the brightness of the galaxy, the same trend does not persist
in the left panel of the second row, where sSFR is shown as a function of surface brightness
and magnitude. For the least luminous galaxies ($M_{B} > -18$), the sSFR is uniformly low, but this is a selection
effect due to the fact that our sample is based on a fixed stellar mass cut, so the 
least luminous galaxies are those with the lowest star formation rates. For more luminous galaxies,
there is clearly a large scatter in the sSFR of galaxies at fixed $\langle \mu_{B} \rangle$
and $M_{B}$.

For $M_{B} < -18$, we find that galaxies with $\langle \mu_{B} \rangle < 25.6$ (HSBGs) have a median
sSFR of $4.3\times10^{-11}$ yr$^{-1}$ while those with $\langle \mu_{B} \rangle > 25.6$ (LSBGs)
have a median value of $5.9\times10^{-11}$ yr$^{-1}$. Although LSBGs have been recorded in 
some papers as having low star formation rates \citep{vanderhulst1993, vandenhoek2000}, others find that their sSFRs 
are not significantly different from those of HSBGs \citep{galaz2011, du2019}, which is similar
to what we find in EAGLE.

The third row of Figure \ref{fig7} shows the mean stellar population age
of each galaxy, computed as the mean age of the star particles
weighted by their initial (i.e. prior to mass loss) particle mass. 
In the right panel, the trend is similar to that seen for sSFR,
such that star-forming galaxies are also younger. Since the subsample of galaxies 
shown in this figure is experiencing a predominantly 
secular evolution, galaxies that are younger tend to
have higher sSFR because they undergo their peak of star formation later in cosmic time \citep{matthee2019}.
However, in the left panel we again see a lack of correlation between
surface brightness and mean stellar age. We find a median stellar age of 7.31 Gyr for
galaxies with $\langle \mu_{B} \rangle < 25.6$ and 7.25 Gyr for those with $\langle \mu_{B} \rangle > 25.6$ ---
a negligible difference.

The bottom panels of Figure \ref{fig7} show the correlation 
of $\langle \mu_{B} \rangle$ and $\langle \mu_{m} \rangle$ with stellar metallicity (total
metal mass fraction) of each galaxy. The mass-metallicity relation (e.g. \citealt{tremonti2004})
is visible on average, but at fixed stellar mass,
there is a significant trend between metallicity and mean stellar
mass surface density. A correlation between metallicity and local stellar density within galaxies has been
noted previously in EAGLE \citep{trayford2019} as well as in 
observations of real galaxies \citep{moran2012, sanchez2013}.
This correlation translates to a somewhat weaker trend with surface brightness in
which lower surface brightness galaxies are more metal-poor, in agreement with
observations of LSBGs \citep{mcgaugh, deblok1998, burkholder2001}.

We now examine the spatial distribution of galaxies as a function of surface brightness.
In the top panel of Figure \ref{fig8}, we show the median (solid lines) and quartiles 
(dashed lines) of the distribution of nearest-neighbor distances for galaxies
split into LSBGs and HSBGs. Here ``neighbors'' are defined as any galaxy with $M_{*} > 10^{9} \mathrm{M_{\odot}}$,
but we see qualitatively similar trends if we use a lower stellar mass limit, or define
``neighbors'' as galaxies with stellar masses above some fraction of the mass of the galaxy being considered.
LSBGs tend to be farther from their nearest neighbor galaxy
than HSBGs, consistent with observations of LSBGs \citep{bothun1993, rosenbaum2004, galaz2011}.

In the second panel of Figure \ref{fig8}, we present the same
plot but for only central galaxies. 
Recall that in EAGLE, every FoF halo has a 
central galaxy that contains the most bound particle,
and all other galaxies are ``satellites''.
Additionally, some centrals have no satellites above
$M_{*} > 10^{9} \mathrm{M_{\odot}}$, meaning their nearest neighbor
is in a different FoF halo. In contrast to the top panel, 
LSBGs and HSBGs have similar distributions of nearest neighbor distances
except at the highest masses. In the third panel of Figure \ref{fig8}, we show statistics for
satellite galaxies only, and here we see a larger difference between LSBGs and HSBGs,
such that HSBG satellite galaxies tend to be closer to their nearest neighbor. The 
difference is more pronounced in the bottom quartile of the nearest neighbor distance
distribution than the top quartile, suggesting that LSBGs are particularly less likely
to have a very close neighboring galaxy.

Finally, in the bottom panel of Figure \ref{fig8}, we show as dashed lines
the fraction of LSBGs and HSBGs that are satellite galaxies as a function
of stellar mass. HSBGs are more likely to be satellite galaxies at all
but the lowest stellar masses, which also contributes to the difference in median 
nearest neighbor distance seen the top panel, as satellite galaxies are on average
closer to their nearest neighbors (note the difference in vertical axis scale
between the second and third panels). However, as dot-dashed lines, we show 
the fraction of LSBG and HSBG satellites with an infall time into their hosts later than 
a cosmic age of 8 Gyr. LSBGs and HSBGs are about
equally likely to be such satellites. This implies that HSBGs are only
more likely than LSBGs to be satellites that were accreted a very long time
ago. 
 
The combination of factors above suggests that LSBGs and HSBGs in EAGLE do not form 
(as central galaxies) in substantially different environments, but rather that encounters with other massive galaxies
tend to disrupt LSBGs. This is why LSBGs are less likely to be satellites, and 
when they are satellites, they are not close-in satellites that fell into their host halo long ago.

\begin{figure}
\begin{center}
  \includegraphics[width=\columnwidth]{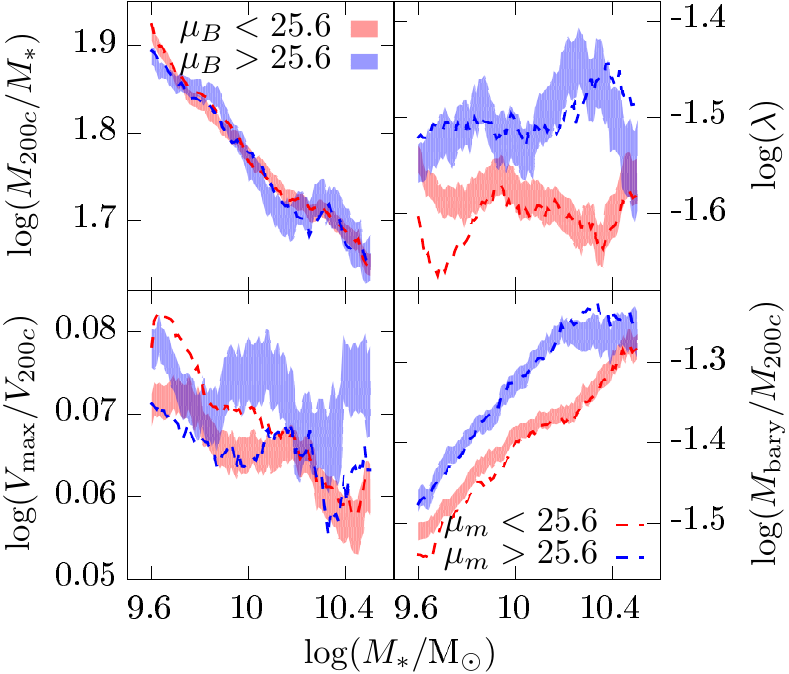}
  \caption{The mean value of various dark matter halo properties as a function of
 central galaxy stellar mass. All halo properties are taken from the 
dark matter only variant of the EAGLE simulation, and are thus
unaffected by baryonic physics (see \S\ref{dmo} for details).
The blue and red bands represent the error around the mean for LSBGs 
($\langle \mu_{B} \rangle > 25.6$ mag/asec$^{2}$)
and HSBGs ($\langle \mu_{B} \rangle < 25.6$ mag/asec$^{2}$), respectively.
Additionally, each panel shows, as dashed lines with the same color coding, 
the mean value for galaxies with low stellar mass surface density 
($\langle \mu_{m} \rangle > 25.6$ mag/asec$^{2}$) and high
stellar mass surface density ($\langle \mu_{m} \rangle < 25.6$ mag/asec$^{2}$). 
\textit{Top left:} The mean ratio of dark matter halo mass ($M_{200c}$) 
to central galaxy stellar mass. LSBGs and HSBGs have essentially
the same stellar-halo mass relation. \textit{Bottom left:}
The mean value of $V_{\mathrm{max}}/V_{200c}$, a proxy for the halo concentration. LSBGs have
slightly more concentrated halos than HSBGs; however, the halo concentration of low \textit{density}
galaxies is slightly \textit{lower} than that of high density galaxies. 
This is because concentration correlates with galaxy age (see text). \textit{Top right:}
The mean halo spin parameter. LSBGs inhabit higher spin halos than HSBGs. \textit{Bottom right:}
The ratio of total baryonic mass (hot gas, cold gas, and stars) in the central subhalo to 
halo mass $M_{200c}$. LSBGs have a higher baryon fraction and are more gas rich than HSBGs.
}
\label{fig10}
\end{center}
\end{figure}

\subsection{Dark matter halo properties of LSBGs with low ex-situ mass fractions} 
\label{resdmo}

Having seen in \S\ref{resnomerge} that LSBGs and HSBGs form in similar environments, we now 
present the correlation of galaxy surface brightness with various 
host dark matter halo properties. While the properties of dark matter halos
are thought to determine many of the properties of the galaxies that
form within them, baryonic
physics can change the distribution of dark matter,
especially on galactic scales, making it challenging to disentangle the influence
of halos on galaxies from that of galaxies on halos. As described in \S\ref{dmo}, we
avoid this problem by examining the properties of ``matched'' dark matter halos
from a dark matter only (DMO) run of EAGLE with identical initial conditions to the Ref-L0100N1504 
reference run. Thus the dark matter halo properties presented here have been
unaffected by baryonic physics. For the reasons given in \S\ref{dmo},
as well as the fact that LSBGs seem to be destroyed by satellite stripping processes,
we present galaxy-halo correlations only for central galaxies/subhalos.

While EAGLE produces, on average, 
realistic galaxy rotation curves \citep{schaller2015},
it has been noted that the scatter in the rotation curves for 
low-mass galaxies is less than what is found in 
observations \citep{oman2015}. LSBGs in particular are 
typically observed to have very slowly rising rotation curves \citep{swaters2000, sparc}.
Factors that have been put forth as possibly contributing
to this discrepancy include the lack of halo core formation in 
EAGLE and similar simulations (\citealt{katz, santos2018}; see however \citealt{benitez2019}),
as well as misestimation of some observed dwarf galaxy rotation
curves due to non-circular gas motions \citep{oman2019}.
The fact that the source of this disagreement has not been found
highlights our incomplete understanding of the relationship between low-mass galaxies
and their dark matter halos. We therefore
caution that it is not certain that EAGLE forms such galaxies
in the correct dark matter halos.
 
In Figure \ref{fig10}, we present the correlation of several halo properties
with the stellar mass of the central galaxy. These centrals are split into
LSBGs and HSBGs using a cut of $\langle \mu_{B} \rangle = 25.6$ mag/asec$^2$. Additionally, the
same correlations are shown for ``low density'' and ``high density''
galaxies, using the same threshold in $\langle \mu_{m} \rangle$. The
curves are computed using a boxcar smoothing of $\pm 0.1$ dex in $M_{*}$.
While the results presented in this figure use the matched DMO
halo properties, we note that similar results are found using
the host halo properties from the reference simulation.

In the top left panel, we show the ratio of the halo mass, $M_{200c}$, 
to the stellar mass, $M_{*}$. It can be seen that LSBGs and HSBGs of the same stellar mass
form in dark matter halos of similar masses. 
In the lower left panel, we show the ratio $V_{\mathrm{max}}/V_{200c}$,
a proxy for the halo concentration \citep{prada2012}. LSBGs 
appear to inhabit slightly more concentrated halos than HSBGs,
but interestingly, this trend seems to
be reversed (at $M_{*} \lesssim 10^{10} \mathrm{M_{\odot}}$) or non-existent 
(at $M_{*} \gtrsim 10^{10} \mathrm{M_{\odot}}$) for galaxies split by their mean surface
density rather than their mean surface brightness. This is likely due to the
fact that halo concentration, which correlates strongly with the assembly
time of the halo \citep{wechsler2002}, has also been found to correlate with 
the mean stellar age of the central galaxy in EAGLE \citep{matthee2017}
and its star formation rate \citep{matthee2019}.
We have seen that galaxies with faint mean surface brightness tend to have
older ages and lower sSFRs relative to galaxies with low surface mass density
(Figure \ref{fig4} and Figure \ref{fig7}). Since the central
galaxies of halos with larger concentrations are older, we
would expect LSBGs to reside in halos with higher concentration than galaxies
selected based on low stellar mass surface density.
 
The top right panel of Figure \ref{fig10} shows the spin parameter, $\lambda$,
of the dark matter halo within $r_{200c}$
as a function of central galaxy stellar mass.
Here there is a noticeable division between LSBGs and HSBGs, such that the former 
are hosted by halos with larger spins. The same is true
for galaxies with low and high stellar mass surface density. However, we note that
while the difference in the median halo spin between
the two populations is statistically significant, it
is much smaller than the scatter in the halo spin within the two groups.
We will comment more on the evolutionary
implications of this in the discussion subsection below.

Finally, in the bottom right panel of Figure \ref{fig10} we show 
the ratio of total baryonic mass in the central subhalo --- including stars and both hot
and cold gas --- to the halo mass, $M_{200c}$. LSBGs have higher baryonic masses
than HSBGs, and given that it is clear from the top left
panel that they do not have substantially higher halo masses at fixed
stellar mass, this implies that they have higher gas masses.
 This is true despite the fact that LSBGs do not 
have significantly higher specific star formation 
rates than HSBGs (Figure \ref{fig6}), meaning that much of their excess
gas mass is in the form of non-star-forming gas. 
%Note, however, that this figure presents the total gas mass rather than 
%specifically the H\textsc{i} mass, which has been found to be higher for
%LSBGs in observations \citep{wyder2009, leisman2017}. Computing the latter
% would require significant postprocessing of the
%simulation outputs (e.g. \citealt{crainhi}), and we do not
%undertake it in this paper.

\subsection{The formation of low-mass LSBGs}
\label{discussion}

Based on the correlations between surface 
brightness and galaxy and halo properties presented above,
a coherent picture emerges of the formation of low surface brightness
galaxies with little mass contribution from mergers, which are the dominant type of LSBG at
stellar masses between $10^{9.5}$ and $10^{10.5} \mathrm{M_{\odot}}$.

First, it is clear from Figure \ref{fig4} and Figure \ref{fig7}
that LSBGs at low stellar masses are to a significant extent the high
angular momentum tail of the galaxy distribution. 
In blue optical bands such as the B band,
there is some scatter in the surface brightness 
at fixed galaxy kinematic morphology that results from scatter
in the mass-to-light ratio, which is highly correlated
with the specific star formation rate of the galaxy and the mean
age of its stellar population. Galaxies that are more
rotation-dominated generally have higher sSFRs, 
but this increases their surface brightness, and thus galaxies
selected to have low surface brightness have some scatter in their kinematic morphology,
and a significant amount in their sSFR and mean stellar age.

The formation of spheroid- and disk-dominated galaxies in EAGLE
has already been examined by a number of authors. 
\citet{zavala2016} found that most stars in disk-dominated
galaxies are formed after the turnaround time of the galaxy's
assembling host dark matter halo. These stars are formed from a gas reservoir
whose angular momentum is set by the spin of the host halo at late times, 
resulting in a correlation between the spin of the stellar component
and that of the host halo. Conversely, the majority of stars
in spheroidal galaxies are formed prior to turnaround, 
and the final angular momentum of the galaxy
is mostly correlated with that of the inner regions of the 
dark matter halo, rather than the entire halo. \citet{zavala2016} attribute
the latter to mergers subsequent to turnaround,
which lead to a loss of angular momentum for both the
stellar component and inner halo.

\citet{clauwens2018} also examined the formation of the spheroid
and disk components of galaxies in EAGLE, finding a ``three-phase'' evolution
set by a galaxy's growth progression through different stellar masses.
In contrast to \citet{zavala2016}, \citet{clauwens2018} find that for $M_{*} \lesssim 10^{10} \mathrm{M_{\odot}}$, 
galaxies grow as kinematic spheroids via
a combination of in-situ star formation and ``tiny'' mergers of
mass ratio less than 1:10. Higher-mass galaxies 
begin to develop disks around their spheroidal component through
in-situ star formation, and at the highest masses ($M_{*} \gtrsim 10^{10.5} \mathrm{M_{\odot}}$), enhancement
of the dispersion-dominated component recommences, but only via mergers. This agrees
with our Figure \ref{fig4}, where it can be seen that
spheroid-dominated galaxies with low ex-situ mass fractions are nearly absent
for $M_{*} > 10^{10} \mathrm{M_{\odot}}$. 
%Similarly, \citet{lagos2016}
%find that galaxies that quench their star formation very early
%can be kinematically spheroid-dominated even in the absence of mergers.
%Presumably, quenching leads such galaxies to be stuck in the ``first phase'' of 
%evolution identified in \citet{clauwens2018}. 

Our results are consistent with the evolutionary scenarios described above.
The galaxies with the lowest surface mass density are those that are 
the most disk dominated. These galaxies are young, 
having formed their stars more recently on average (Figure \ref{fig7}),
and they inhabit halos with higher spins (Figure \ref{fig10}), in agreement
with \citet{zavala2016}. However, low surface brightness
is anticorrelated with young stellar population ages and high star formation rates 
(Figure \ref{fig3} and Figure \ref{fig7}), so there is no
remaining correlation between galaxy surface brightness and stellar population age.
Nevertheless, galaxies with the highest
angular momentum are still those in the highest-spin halos,
and thus a significant correlation remains between host halo
spin and galaxy surface brightness.

In addition to the properties presented in Figure \ref{fig10}, we also
computed the correlation of the surface brightness with the spin
of the ``inner halo'', defined as the particles within $0.1r_{200c}$.
While we did find a correlation, it was lower than that
between the surface brightness and the spin of the entire halo.
This is in agreement with the conclusions of \citet{zavala2016},
which imply that low-mass galaxies with few mergers 
should have angular momentum that is better 
correlated with the large-scale spin of the halo.

Our conclusions agree partially with the recent work of \citet{dicintio2019}
using the NIHAO suite of zoom-in hydrodynamical simulations. 
The authors found that LSBGs with masses 
$10^{9.5} < M_{*}/\mathrm{M_{\odot}} < 10^{10}$ form in halos with higher spins, 
and lack a significant correlation with any other halo parameters. 
However, they also found that this
is primarily the result of galaxy mergers, with higher spin
galaxies and halos having had more aligned rather than misaligned mergers.
Here we find that galaxies that have had very little merger
activity have a significant range of kinematic morphologies and surface
brightnesses. Furthermore, the amount of gas present in the halo
appears to play a significant role in the formation of extended
disks (Figure \ref{fig10}). 

\subsection{The effect of mergers on surface brightness}
\label{resmergers}

In the previous subsection, we described how low surface brightness galaxies
can form through (nearly) secular evolution, via growth of their stellar
disks at late times from a reservoir of gas co-rotating with a high-spin host halo. However, the 
highest-mass galaxies in EAGLE, with stellar masses comparable to the estimated
value for Malin 1 ($\approx 10^{11} \mathrm{M_{\odot}}$; \citealt{boissier2016}), have all
undergone significant mass growth from mergers (Figure \ref{fig4}).

From Figure \ref{fig4} and Figure \ref{fig5}, it is clear that
mergers typically make galaxies more kinematically dispersion dominated,
and also build up a faint outer stellar halo. However, in Figure \ref{fig4}
it is also possible to see a few massive galaxies with significant
mass growth from mergers that nevertheless have $\bar{\epsilon_{*}}(R_{28B}) > 0.5$,
implying that they are disk-like in their outer regions.
For the remainder of this subsection,
we will focus on the role of mergers in creating large, low-density disks.

\begin{figure}
  \begin{minipage}{0.5\columnwidth}
    \centering
    \includegraphics[width=.95\columnwidth]{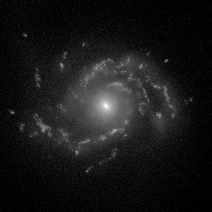} 
    \vspace{1ex}
  \end{minipage}%%
  \begin{minipage}{0.5\columnwidth}
    \centering
    \includegraphics[width=.95\columnwidth]{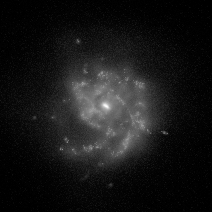} 
    \vspace{1ex}
  \end{minipage} 
  \begin{minipage}{0.5\columnwidth}
    \centering
    \includegraphics[width=.95\columnwidth]{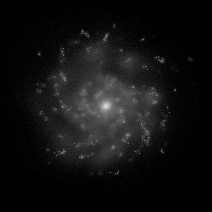} 
    \vspace{1ex}
  \end{minipage}%% 
  \begin{minipage}{0.5\columnwidth}
    \centering
    \includegraphics[width=.95\columnwidth]{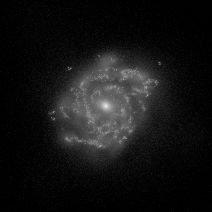} 
    \vspace{1ex}
  \end{minipage}
\caption{Synthetic B-band face-on images (including dust) of 
the four galaxies possessing an outer disk ($\bar{\epsilon_{*}}(R_{28B}) > 0.5$)
having the largest values of $R_{28B}$ in our sample.
Images were created using the radiative transfer code SKIRT (see Appendix \ref{appendix}),
and are 200 pkpc per side with pixel size 0.94 pkpc
and logarithmic flux scaling. \textit{Top Left:} GalaxyID
15548147, $M_{*} = 1.2\times10^{11} \mathrm{M_{\odot}}$, $R_{28B} = 62.2$ kpc, $\langle \mu_{B} \rangle = 26.1$
mag/asec$^{2}$. \textit{Top Right:} GalaxyID
16643441, $M_{*} = 8.1\times10^{10} \mathrm{M_{\odot}}$, $R_{28B} = 56.1$ kpc, $\langle \mu_{B} \rangle = 25.7$
mag/asec$^{2}$. \textit{Bottom Left:} GalaxyID
17668706, $M_{*} = 4.2\times10^{10} \mathrm{M_{\odot}}$, $R_{28B} = 55.5$ kpc, $\langle \mu_{B} \rangle = 26.4$
mag/asec$^{2}$. \textit{Bottom Right:} GalaxyID
16169302, $M_{*} = 5.6\times10^{10} \mathrm{M_{\odot}}$, $R_{28B} = 53.3$ kpc, $\langle \mu_{B} \rangle = 25.7$
mag/asec$^{2}$.}
  \label{figm2} 
\end{figure}

\begin{figure*}
  \begin{center}
    \includegraphics[width=\textwidth]{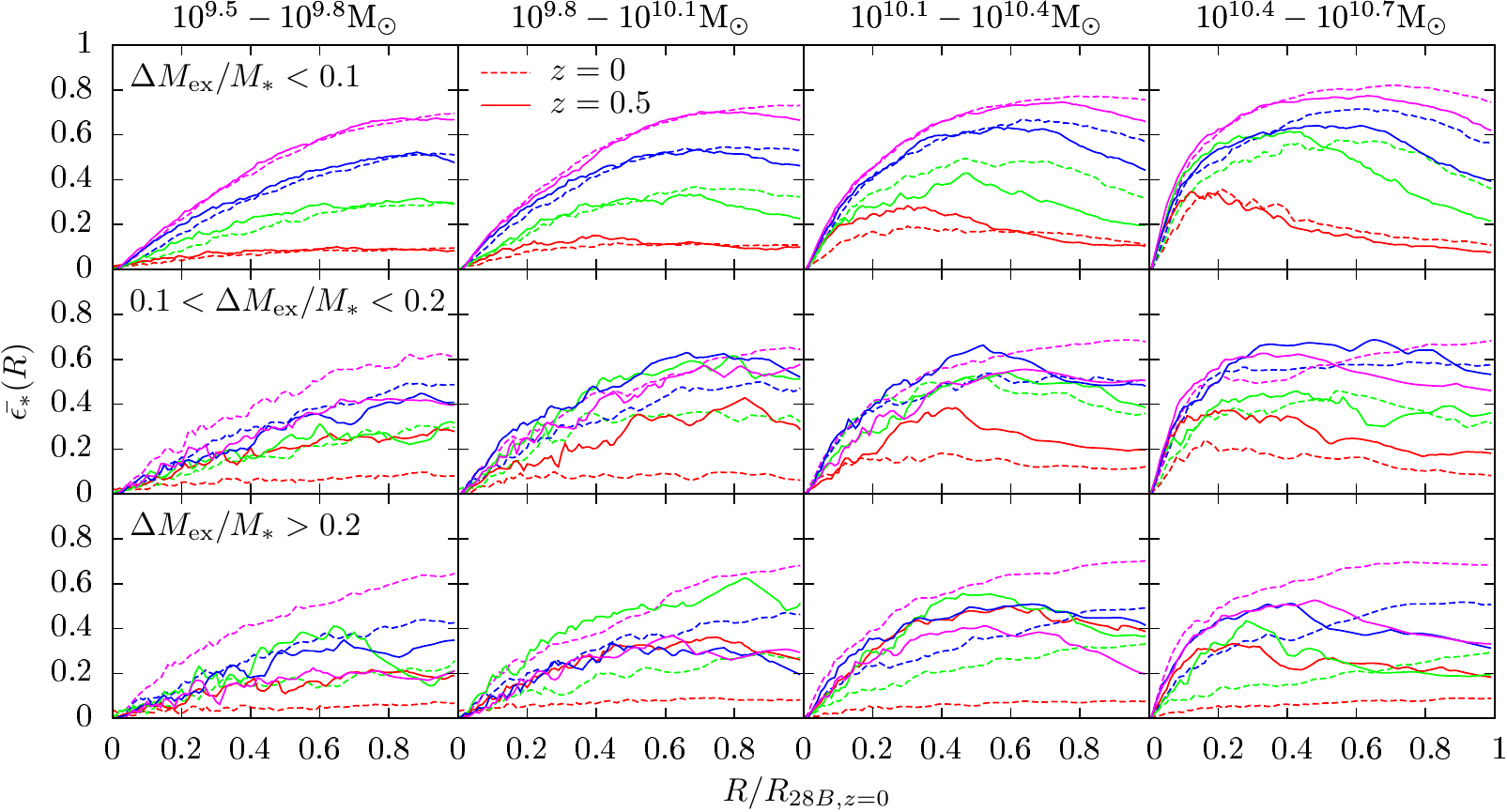}
    \caption{The evolution of $\bar{\epsilon_{*}}(R)$ ($\epsilon_{*}$ as a function
of projected radius) between $z = 0.5$ and $z = 0$ for galaxies with different properties. 
The columns split galaxies into bins of $z = 0$ stellar mass.
The rows show bins in $\Delta M_{\mathrm{ex}}/M_{*}$, the increase 
in ex-situ stellar mass between $z = 0.5$ and $z = 0$, expressed
as a fraction of the total stellar mass at $z = 0$. Galaxies that
have lower stellar mass at $z = 0$ than $z = 0.5$ (generally due to stripping while satellites)
are excluded from the bins.
A limited range of galaxy stellar masses are shown ($M_{*} < 10^{10.7} \mathrm{M_{\odot}}$) in order
to select masses at which galaxies exist with a variety of $\Delta M_{\mathrm{ex}}/M_{*}$.
In each panel, galaxies are binned into four different ranges of $\bar{\epsilon_{*}}(R_{28B})$ at $z = 0$:
$0 - 0.2$ (red), $0.2-0.4$ (green), $0.4-0.6$ (blue) and 
$0.6-0.8$ (magenta). Note that different bins of $M_{*}$ and $\Delta M_{\mathrm{ex}}/M_{*}$
contain significantly different distributions of $\bar{\epsilon_{*}}(R_{28B})$.
Dashed lines represent the median 
$\bar{\epsilon_{*}}(R)$ for each bin as a function of $R/R_{28B}$ at $z = 0$.
Using the same $z = 0$ physical $R_{28B}$ to 
scale the radius at $z = 0.5$, the median $\bar{\epsilon_{*}}(R)$ curve for the main progenitors of the galaxies in each
bin is plotted versus $R/R_{28B}$ as solid lines. For galaxies with little
contribution from mergers, the relative ordering of $\bar{\epsilon_{*}}(R_{28B})$ is 
effectively fixed before $z = 0.5$, whereas for galaxies with a significant merger contribution,
there is far more change in the kinematic morphology, with some galaxies becoming more 
dispersion-dominated but others becoming more rotation-dominated in their outer regions.} 
    \label{figm3}
  \end{center}
\end{figure*}

\begin{figure*}
  \begin{center}
    \includegraphics[width=\textwidth]{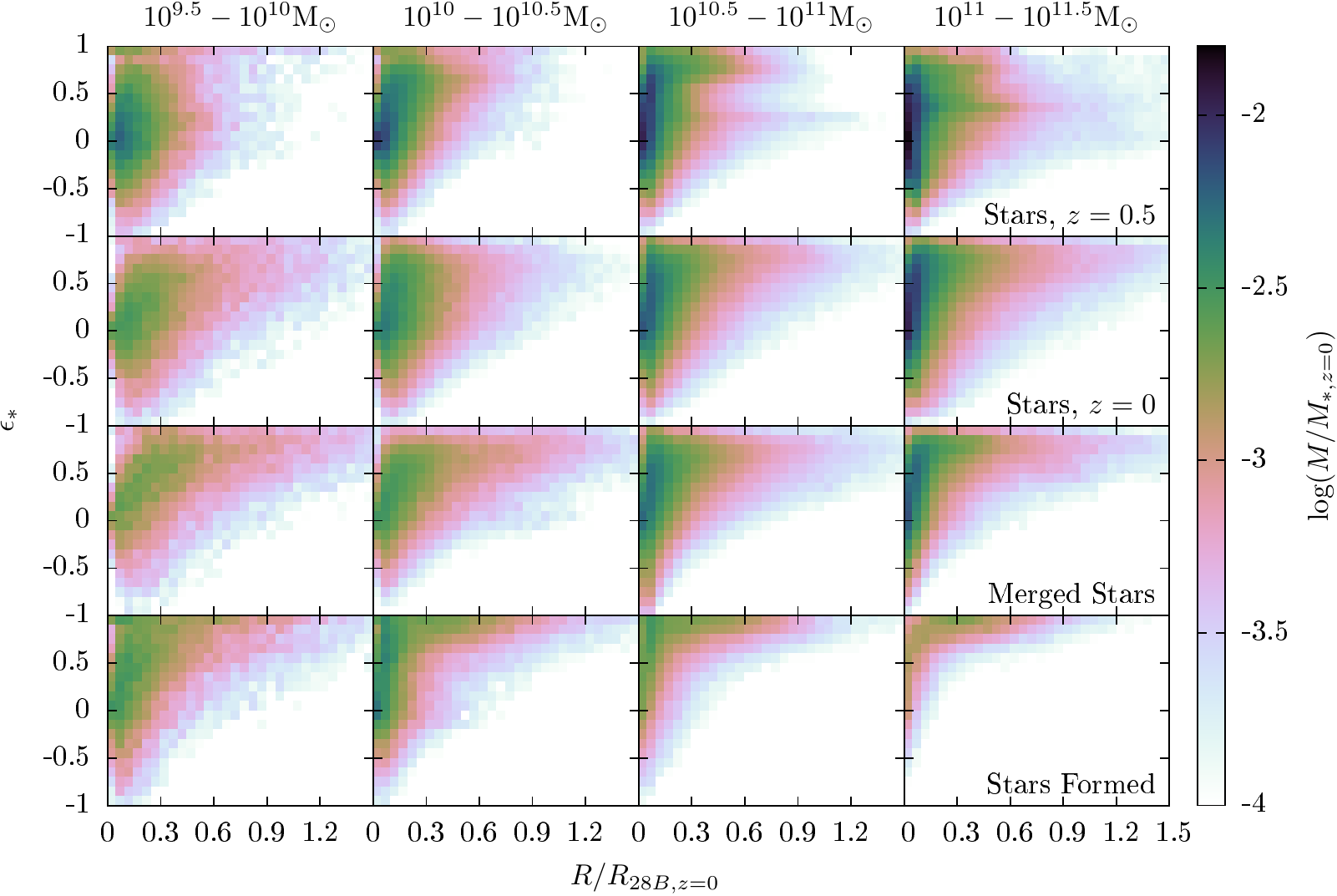}
    \caption{Mean star particle distribution in projected radius $R$ and orbital
circularity parameter $\epsilon_{*}$ for galaxies with $\bar{\epsilon_{*}}(R_{28B}) > 0.5$
at $z = 0$ that gained $>25\%$ of their $z = 0$ stellar mass from ex-situ stars since $z = 0.5$. Different
columns show different bins of $z = 0$ stellar mass. The rows show the distributions
of different subsets of stars. All panels show the projected 
radius scaled by  $R_{28B}$ at $z = 0$, with color coding indicating
the stellar mass in each bin as a fraction of the $z = 0$ stellar mass of the galaxy.
\textit{Top row:} The distribution of the stars
in the main progenitor galaxy at $z = 0.5$. \textit{Second row:} The
same stellar particles as in the top row, but now their distribution 
in the descendant galaxy at $z = 0$. Compared to $z = 0.5$, the 
distribution of stellar particles becomes more
kinematically disk-dominated.
\textit{Third row:} The ex-situ stars accreted
from other galaxies between $z = 0.5$ and $z = 0$. These stars contribute
significantly to the outer disk of the galaxy. 
\textit{Bottom row:} The stars that formed between $z = 0.5$ and $z = 0$ in
the main progenitor. Star formation also contributes to the outer disk, especially
for lower-mass galaxies.}
    \label{figm4}
  \end{center}
\end{figure*}

\begin{figure*}
  \begin{center}
    \includegraphics[width=\textwidth]{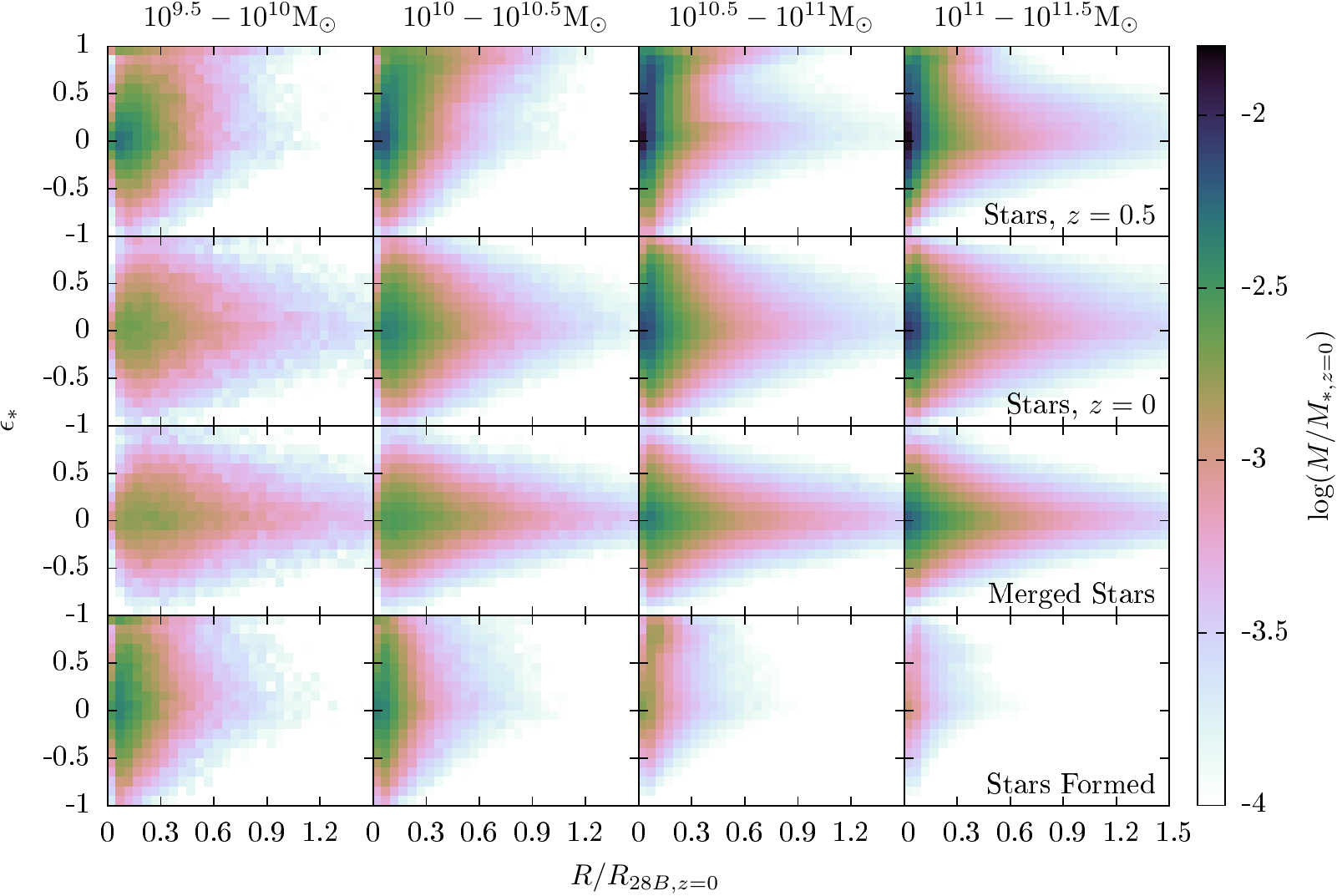}
    \caption{The same as Figure \ref{figm4}, but now for galaxies with $\bar{\epsilon_{*}}(R_{28B}) < 0.2$
(dispersion-dominated) at $z = 0$ in addition to $\Delta M_{\mathrm{ex-situ}}/M_{*}(z = 0) > 0.25$. A 
number of differences in the star particle distribution are visible in comparison to
Figure \ref{figm4}. The stars present in the main progenitor at $z = 0.5$ become
more kinematically spheroid-dominated by $z = 0$. Stars acquired via mergers contribute to a kinematically
hot stellar halo in the outer parts of the galaxy rather than a disk, 
while newly-formed stars do not contribute to the outer parts of
the galaxy at all.} 
    \label{figm5}
  \end{center}
\end{figure*}

To select the most extended galaxies whose faint outer regions are disk-like
rather than a dispersion-dominated stellar halo, 
we select all the galaxies that have $\bar{\epsilon_{*}}(R_{28B}) > 0.5$
and examine the ones with largest $R_{28B}$. On visual inspection, the largest such
galaxy turns out not to be a disk galaxy but a ring
galaxy. (Ring galaxies in EAGLE were studied in \citealt{elagali2018}.)
We thus exclude it from consideration. For the four next
largest galaxies, we show mock face-on B-band images in Figure \ref{figm2}.
These images were created using the Monte Carlo radiative transfer code
SKIRT \citep{skirt2003, skirt2011, skirt2015}, as described in Appendix \ref{appendix}.
Note that while we neglect the effects of dust when computing
surface brightnesses in this paper, these images
include a model for dust extinction.
Spiral structure can be seen surrounding the nucleus of each galaxy in
Figure \ref{figm2}, consistent with the morphology
inferred from $\bar{\epsilon_{*}}(R_{28B})$.

The largest disk galaxy in our sample has $R_{28B} = 62$ kpc.
For comparison, $R_{28B}$ of Malin 1 is $\approx 80$ kpc \citep{boissier2016}, while
$R_{28B}$ of UGC1382 is approximately equal to
that of our largest galaxy \citep{hagen2016}. Thus the largest
galaxies in our sample have sizes comparable to those of observed giant LSBGs. 

The ex-situ stellar mass fractions of the four galaxies in 
Figure \ref{figm2} are 0.57, 0.19, 0.27, and 0.33, in order
of decreasing $R_{28B}$, meaning that all four of these galaxies have 
experienced a significant mass contribution from mergers.
In fact, the largest 17 disk galaxies in our sample have $f_{\mathrm{ex-situ}} > 0.1$.
However, the largest galaxy with $f_{\mathrm{ex-situ}} < 0.1$ has 
$R_{28B} = 48$ kpc, so the gap between the sizes of the largest
disk galaxies that had a mostly secular evolution and those with substantial
mergers is not very large. This may be due to statistics: the box size of
EAGLE is not sufficient to host large numbers of massive galaxies with
many mergers, so it is unlikely to contain the most unusually large galaxies
that can be found in our Universe.

Thus, despite the general tendency of mergers
to make galaxies less disk-dominated, some galaxies 
with significant mass growth from mergers have 
large extended disks. To examine in more detail
the range of possible outcomes of recent mergers, 
we show in Figure \ref{figm3} the kinematic evolution
of galaxies with different stellar mass contribution from mergers
between $z = 0.5$ and $z = 0$. For different bins of $\bar{\epsilon_{*}}(R_{28B})$
at $z = 0$, we show $\bar{\epsilon_{*}}$ as a function of radius for the $z = 0$
galaxies and their $z = 0.5$ main progenitors. Galaxies
with lower stellar masses at $z = 0$ than $z = 0.5$ have
been excluded from the figure as they are generally
undergoing stripping as satellites, and this process
is likely to affect their kinematic morphology in a unique way. We 
have also seen in the previous two subsections that LSBGs are unlikely to survive
being stripped. We therefore exclude this subset of galaxies from our
sample in the remainder of this subsection.

It is apparent in Figure \ref{figm3} that galaxies with 
little contribution from mergers have their relative ordering
in kinematic morphology largely fixed by $z = 0.5$. Their evolution 
in $\bar{\epsilon_{*}}(R)$ varies with galaxy stellar mass. 
Galaxies with $M_{*} < 10^{9.8}$ at $z = 0$ have not changed
their kinematic profiles significantly since $z = 0.5$, whereas for
higher $M_{*}$, the buildup of rotationally-supported disks can be seen, consistent
with the evolutionary picture presented in \citet{clauwens2018}.
Specifically, galaxies become
more kinematically disk dominated in their outer parts near $R_{28B}$.

Looking at the panels in Figure \ref{figm3} that represent galaxies
with a higher merger contribution between
$z = 0.5$ and $z = 0$, we see that the correlation between
kinematic morphology at these two redshifts becomes weaker as mergers
become more important, especially for the lowest-mass galaxies.
We note that galaxies are not evenly distributed in number
between the different $\bar{\epsilon_{*}}(R_{28B})$ bins, as galaxies
with large mass contributions from mergers are predominantly
dispersion-dominated. The majority of such
galaxies follow the trend shown by the red lines in the bottom panels
of Figure \ref{figm3}, decreasing in $\bar{\epsilon_{*}}$ with time over
their full radial extent, and especially at large radii. 
However, for the highest bin of $\bar{\epsilon_{*}}(R_{28B})$
at $z = 0$, shown in magenta, galaxies are more rotation-dominated 
in their outer parts at $z = 0$ than at $z = 0.5$, despite undergoing
substantial mergers. This suggests that mergers might in some cases
extend the disks of galaxies rather than destroying them.

To explore this idea further, we select galaxies with 
a significant contribution to their stellar mass from
mergers between $z = 0.5$ and $z = 0$, such that 
at least $25\%$ of their $z = 0$ stellar mass was found
within other subhalos after $z = 0.5$ but before $z = 0$
(i.e. $\Delta M_{\mathrm{ex-situ}}/M_{*}(z = 0) > 0.25$). We then
compute the two-dimensional distribution of their stellar
particles in projected radius $R$ and orbital circularity $\epsilon_{*}$,
to see how the particles are distributed physically
and kinematically. We do this for three different subsets
of the $z = 0$ stellar particles: those that were
already in the main progenitor galaxy at $z = 0.5$, those that were
accreted from other galaxies between $z = 0.5$
and $z = 0$, and those that formed within the
main progenitor between $z = 0.5$
and $z = 0$. We also show the original distribution of the star particles 
in the $z = 0.5$ main progenitor galaxy for comparison.

We compute the aforementioned distributions for galaxies 
that have extended disks ($\bar{\epsilon_{*}}(R_{28B}) > 0.5$)
and those that do not ($\bar{\epsilon_{*}}(R_{28B}) < 0.2$),
as well as four different $z = 0$ stellar mass bins between $10^{9.5} \mathrm{M_{\odot}}$ and 
$10^{11.5} \mathrm{M_{\odot}}$. We then obtain the mean fraction 
of the total $z = 0$ $M_{*}$ contained in each $(R/R_{28B}, \epsilon_{*})$ bin.
The particle distributions for the 
rotation-dominated and dispersion-dominated galaxies are
shown in Figure \ref{figm4} and Figure \ref{figm5}, respectively.

\begin{figure}
  \begin{center}
    \includegraphics[width=\columnwidth]{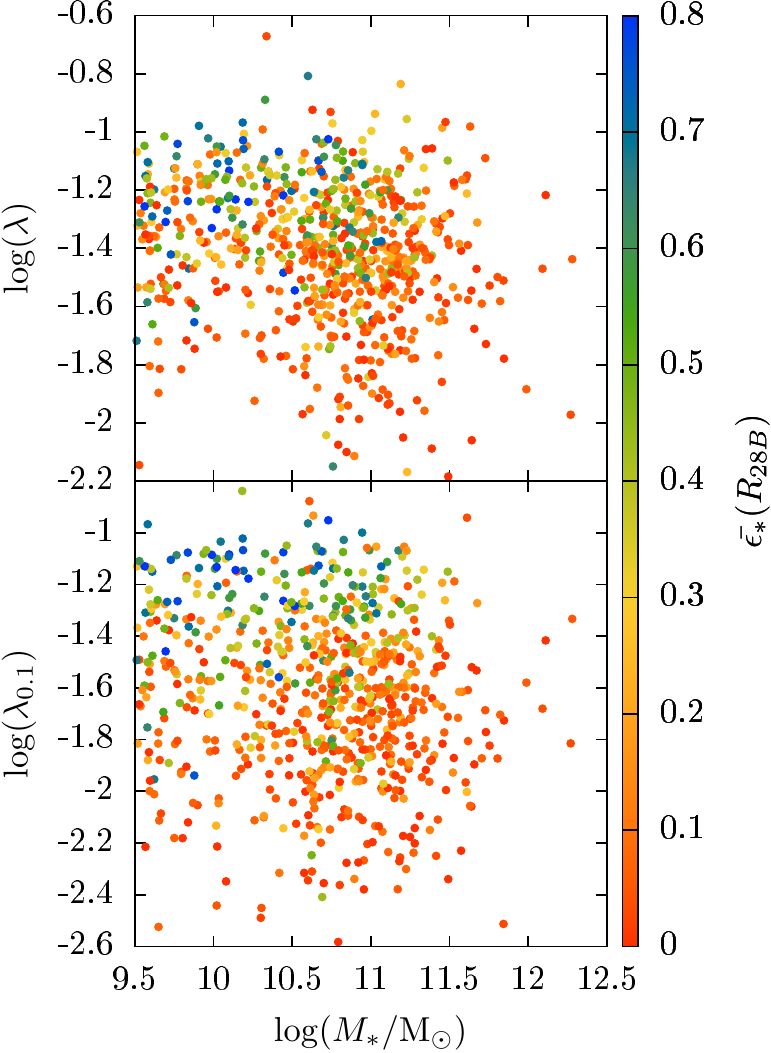}
    \caption{For all central galaxies with $f_{\mathrm{ex-situ}} > 0.3$, the galaxy stellar mass ($M_{*}$) versus
the spin ($\lambda$) of the dark matter halo in the dark matter only EAGLE simulation
that corresponds to the galaxy's host halo (see \S\ref{dmo} for details). The top panel shows the spin parameter of the entire halo 
(within $r_{200c}$) while the lower panel shows the spin only of the inner halo (within $0.1r_{200c}$).
Color coding in both panels indicates the local orbital circularity, $\bar{\epsilon_{*}}$, at $R_{28B}$. The orbital
circularity correlates with the total halo spin, and even more so with the inner halo spin.} 
    \label{figm6}
  \end{center}
\end{figure}

Since the color coding shows the stellar mass in each bin as a fraction
of the total $z = 0$ stellar mass of the galaxy, we see that
for both dispersion- and rotation-dominated galaxies, the fraction
of stellar mass from star formation between $z = 0.5$ and $z = 0$ 
(bottom row) is higher at lower stellar masses, decreasing
from $33\%$ in the lowest mass bin to $\le 11\%$ in the highest. 
At fixed stellar mass, however, 
there is only a small difference between
disk and spheroidal galaxies in the fraction of mass gained
from star formation. The largest difference occurs in the highest-mass bin, where disk galaxies
obtain an average of $11.0\%$ of their $z = 0$ mass from star formation,
while for spheroids the fraction is $7.1\%$. 
However, the distributions of this value are substantially overlapping:
the one-sigma scatter is $6.2\%$ for disk galaxies and 
$4.1\%$ for spheroids. The difference between spheroid and disk galaxies decreases
towards lower masses, such that in the lowest mass bin,
the two types of galaxies experience an equal amount of
star formation on average.

More significantly different is how the recently-formed stars are distributed in projected radius
and orbital circularity. For spheroidal galaxies, the stars form 
overwhelmingly in the inner regions of the galaxy ($R/R_{28B} < 0.3$),
whereas for disk galaxies there is a tail of
stars on very cold orbits that extends to $\gtrsim R_{28B}$ 
visible at the top of the bottom panels in Figure \ref{figm4}.

We split the stars formed in the galaxy into different subgroups
depending on whether their parent gas particles entered
the main progenitor via mergers or accretion,
as well as the redshift at which they became part of the main progenitor.
However, we found that the source of the gas
does not significantly affect the distribution of the resulting
stars in $R$ and $\epsilon_{*}$, nor does the 
redshift at which the gas became bound.

In the third row of panels in Figure \ref{figm4}
and Figure \ref{figm5}, we see the distribution of star particles
that entered the galaxy via mergers between $z = 0.5$
and $z = 0$. (The fraction of stellar mass from
mergers has been chosen to be the same for both disk
and spheroidal galaxies.) Again, the distribution of these stars
is highly different for the selected disk and spheroidal
galaxies. For the spheroidal galaxies at $z = 0$, the
ex-situ stars are kinematically hot regardless of radius.
They predominantly come to inhabit the inner regions of the galaxy,
but also form a long tail to $R_{28B}$ and beyond. By comparing
the radial distribution of these stars to those in the panels
above (second row), which show the $z = 0$ distribution of stars that
were already present in the galaxy at $z = 0.5$, we see that
the ex-situ stars at large $R$ are building up the stellar halo, 
especially in the lower-mass bins. This is consistent
with the surface density profiles presented
in Figure \ref{fig5}.

Conversely, for the galaxies that are kinematically disk-dominated
at $z = 0$, the distribution of ex-situ stars 
that extends to large $R$ has high orbital circularity, although
not quite as high as the stars formed since $z = 0.5$. Nevertheless,
the mass in stars gained directly from mergers is comparable to or larger
than the mass in stars formed since $z = 0.5$ for all the galaxy mass bins
presented, implying that ex-situ stars are a large
contribution to the highly extended disk seen in our largest LSBGs.

Finally, we examine the top two rows of Figure \ref{figm4}
and Figure \ref{figm5}, showing the distribution at $z = 0.5$
and $z = 0$ of the star
particles that have been bound to the galaxies since $z = 0.5$.
Because we choose the galaxies presented in the two figures solely based on their $z = 0$
value of $\bar{\epsilon_{*}} (R_{28B})$, 
the $z = 0.5$ progenitors of the disk-dominated galaxies are
already somewhat more kinematically cold in their outer parts than
the progenitors of the spheroidal galaxies (top row). However, the evolution
of these star particles increases this difference. For the 
$z = 0$ spheroidal galaxies, the $z = 0.5$ progenitors
initially have a cold rotating component in all mass bins, visible in
the top row of panels in Figure \ref{figm5}. However,
by $z = 0$, this cold component has essentially disappeared, as seen
in the second row of panels. This is the ``typical''
effect of mergers: to make galaxies more spheroid-dominated.
This can be also be inferred from the distribution of 
morphologies as a function of $f_{\mathrm{ex-situ}}$ in Figure \ref{fig4}.
However, for the $z = 0$ disk galaxies in Figure \ref{figm4}, 
the cold component is enhanced in all mass bins.

Thus we can conclude that while mergers do typically transform
galaxies into spheroids, there is a subset of mergers that instead
builds up a rotationally-supported, extended disk, perhaps
leading to the formation of a giant LSBG.

\subsection{Comparison to prior work on kinematic evolution due to mergers in EAGLE}
\label{lagos}

The effect of mergers on the kinematics of galaxies in EAGLE has already been
studied in \citet{lagos2018}. The authors found that different types
of mergers impact the specific angular momentum of galaxies differently. Dry mergers
were found to significantly decrease the specific angular momentum of galaxies,
whereas wet mergers on average slightly increased it. Moreover,
particular types of wet mergers were found to be more likely to spin
up the galaxy; these include minor (as opposed to major) mergers, mergers in which
the satellite galaxy's spin is aligned with that of the central, and 
mergers in which the merging satellite has a large orbital angular momentum.

In \citet{lagos2018}, ``wet'' mergers were defined based on the 
neutral gas fraction, which we do not compute in this paper, thus precluding
a direct comparison. We do note that \citet{lagos2018} find
a larger increase in stellar mass surface density
$\approx 1$ Gyr after wet mergers than dry ones.
This may be consistent with the difference 
in stellar mass gain from star formation between our kinematic
disk and spheroid galaxy samples with $M_{*} > 10^{10.5} \mathrm{M_{\odot}}$.
However, at lower stellar masses, there is little difference between the mass
in stars formed since $z = 0.5$ in disk and spheroid galaxies. Thus,
the influence of wet versus dry mergers is unlikely to fully explain
the different angular momentum evolution
seen in Figures \ref{figm4} and \ref{figm5}.

Throughout our paper, we consider any star particles that were once
bound to a subhalo other than the main progenitor to be ``ex-situ'' particles.
\citet{lagos2018}, however, define mergers as only those events
in which the mass ratio of the two galaxies is at least 1:10,
and any smaller mergers are considered ``accretion''. They
also find that accretion tends to increase the angular momentum of galaxies.
To check if this affects our results, we broke down the ex-situ contribution to the 
galaxies in our sample based on the fraction of ex-situ mass from individual
galaxies (i.e., a merger with a more massive galaxy results in more
ex-situ stars from that galaxy). We do not find strong differences
in the distribution of stellar masses contributed by single objects
between the disk and spheroidal subsamples. Therefore, differing mass 
contributions from accretion, minor, and major 
mergers are also unlikely to be the main reason for the buildup
of disks subsequent to mergers in our sample.

Additionally, \citet{lagos2018} find an increase in the angular momentum of 
the outer regions of galaxies (to $\approx 30$ kpc) following wet minor mergers with large
orbital angular momentum. However, they find that this difference comes
from the formation of new stars with high angular momentum at large radii,
but not from ex-situ stars gained during the merger. 
Thus our results appear to depart somewhat from those of 
\citet{lagos2018}, as we find that both newly-formed stars and ex-situ
stars from mergers can contribute to the formation of a large rotationally
supported disk. 

It is possible that our results differ simply because we are
selecting a very unusual subsample of galaxies, whereas
\citet{lagos2018} consider all galaxies that experience mergers. Galaxies with
a large fraction of ex-situ stars that also have large disks
are rare, as can be seen in Figure \ref{fig4}.

One relevant correlation is presented in Figure \ref{figm6}, which
plots galaxy stellar mass versus dark matter halo spin parameter,
color-coded by $\bar{\epsilon_{*}} (R_{28B})$, for 
central galaxies with $f_{\mathrm{ex-situ}} > 0.3$. In the upper panel,
the spin parameter of the halo is computed within $r_{200c}$,
while in the lower one it is computed within $0.1r_{200c}$. As in Figure \ref{fig10}, 
the halo properties are those of galaxies' matched host halos
in the dark matter only (DMO) run of EAGLE, so they
are unaffected by the baryonic physics present in the reference
simulation. In the top panel
of Figure \ref{figm6}, we see that there exists a correlation between large-scale
halo spin and galaxy kinematic morphology. In fact, this correlation is
stronger than the one seen for galaxies with low $f_{\mathrm{ex-situ}}$
in Figure \ref{fig10}. However, looking at the lower panel,
the correlation with the spin of the inner halo is even stronger, such
that nearly all galaxies with $\bar{\epsilon_{*}} (R_{28B}) > 0.5$ have
matched DMO halos with the highest inner spin parameters. This is consistent
with the results of \citet{zavala2016}, since galaxies that have
gained a significant fraction of their mass from mergers are those
that formed most of their stars before their halo turnaround time,
and according to \citet{zavala2016} their angular momentum
should correlate better with that of the inner dark matter halo. 

Given that the DMO halo spins result entirely from the dark matter
initial conditions (which determine the accretion to the halo
and mergers with other halos over time), the strong correlation
between the galaxy angular momentum and the inner halo spin suggests that processes
unrelated to baryonic physics are the primary drivers of the angular momentum evolution
of galaxies that undergo many mergers. This might also indicate that galaxies that grow their disks from mergers
live in unusual large-scale environments. However, we find further
exploration of this topic to be beyond the scope of this work.

\section{Conclusions}
\label{conclusions}

We examined the formation and evolution of low surface brightness galaxies (LSBGs)
and how they differ from high surface brightness galaxies (HSBGs) within the (100 cMpc)$^{3}$ 
reference run of the EAGLE suite of hydrodynamical cosmological simulations. 
We computed synthetic B-band surface brightness profiles
for galaxies with $M_{*} > 10^{9.5} \mathrm{M_{\odot}}$ using the
EAGLE catalog dust-free $ugriz$ luminosities \citep{trayford2015}
of all the star particles bound to each subhalo.
We took all galaxies to be in the face-on orientation in order to maximize the number of objects that
could potentially be observed as LSBGs if they existed in our Universe.

In order to be able to identify nucleated LSBGs, which are bright in 
their central regions, we parametrize the ``effective'' surface
brightness via the 28 mag/asec$^{2}$ B-band isophote, $R_{28B}$, and the mean B-band
surface brightness within it, $\langle\mu_{B}\rangle$. For the majority of low-mass galaxies
($M_{*} \lesssim 10^{10.5}$), $\langle\mu_{B}\rangle$ separates galaxies into LSBG and HSBG subgroups. For this subpopulation,
the median surface brightness of galaxies becomes brighter with increasing stellar mass, as expected. However, galaxies
that have had a large mass contribution from mergers (which includes essentially all
galaxies with $M_{*} \gtrsim 10^{11}$) build up a faint stellar halo which increases
the size of the 28 mag/asec$^{2}$ isophote and thus decreases the mean surface brightness
 within it (Figure \ref{fig5}). Because of this and the fact that undergoing mergers can
significantly alter the evolution of a galaxy, we investigate separately
the formation of LSBGs through secular evolution 
(which is the case for most low-mass LSBGs) and the growth of LSBG disks
via mergers (which produces a small number of giant LSBGs). Our main conclusions are as follows:
\begin{itemize}
\item Surface brightness in the B band is
determined by a combination of the mean surface density of the galaxy, which is related to its physical evolution,
and its specific star formation rate, which determines its mass-to-light ratio and 
can vary on short timescales \citep{matthee2019}. This is shown
in Figure \ref{fig3}. The correlation between higher sSFR and
higher surface brightness introduces scatter into the correlations between surface brightness
and galaxy physical properties that are tightly correlated with stellar mass surface density (Figure \ref{fig7}).
\end{itemize}
For galaxies that experience predominantly secular growth, 
which includes the majority of galaxies with $M_{*} < 10^{10.5} \mathrm{M_{\odot}}$:
\begin{itemize}
\item The mean stellar mass surface density is almost entirely determined 
by the kinematic morphology and the stellar mass
of the galaxy. This can be seen in Figure \ref{fig4} and Figure \ref{fig7}.
This leads to LSBGs generally being highly kinematically disk dominated.
\item LSBGs are on average farther from their nearest neighbor galaxy
 than HSBGs, in agreement with observations. However, we find
that this trend is driven entirely by the fact that LSBGs are unlikely to be
close-in satellite galaxies (Figure \ref{fig8}). This implies that LSBGs do not form
in unusually isolated environments, but rather that they are destroyed by close
encounters with massive galaxies.
\item LSBGs that are the central galaxy in their
host dark matter halo inhabit halos
with similar masses and concentrations as those
of central HSBGs, although LSBGs tend to have a higher baryon fraction 
(including both hot and cold gas; Figure \ref{fig10}).
\item LSBGs whose evolution was not strongly influenced by mergers appear
to form by a combination of the evolutionary processes presented in 
 \citet{zavala2016} and \citet{clauwens2018}.
In this picture of secular galaxy evolution, spheroids form at early times, and disks form around
them later from a reservoir of gas that co-rotates with the host dark matter halo.
Thus galaxies with high angular momentum tend to be those that formed most 
of their stars recently, as well as those with higher dark matter halo spins. The correlation
between surface brightness and sSFR means that selecting LSBGs selects against
young galaxies (Figure \ref{fig7}); however, the correlation between
low surface brightness and high dark matter halo spin remains (Figure \ref{fig10}).
\end{itemize}
Regarding the influence of mergers on galaxy surface
brightness profiles and morphologies:
\begin{itemize}
\item Mergers generally make galaxies more spheroidal.
However, a small subset of galaxies with large
ex-situ fractions are kinematically disk-dominated in their faint outer
regions (Figure \ref{fig4}). This subset represents some of the largest
LSBGs in our sample, such as those shown in Figure \ref{figm2}.
\item The ex-situ stars gained from mergers generally build up a faint
dispersion-dominated stellar halo component that extends to large radii,
 as can be seen in the third row of Figure \ref{figm5}. However, in a minority of cases, 
the stars from mergers can instead build up extended disks, as seen in Figure \ref{figm4}.
Merger-related growth is perhaps necessary for the formation of the largest
``giant'' LSBGs seen in the Universe.
\item For galaxies with a high ex-situ stellar mass fraction, the
kinematic morphology of their faint outer regions
correlates strongly with the spin of the inner part ($<0.1r_{200c}$) 
of their matched dark matter halo from the dark matter only EAGLE run (Figure \ref{figm6}). 
Given that the dark matter only simulation
is unaffected by baryonic physics, this implies that galaxies
that grow their disks via mergers may be located in an unusual
dark matter environment.
\end{itemize}

It is possible that we have not identified every 
mechanism by which low surface brightness galaxies are able to form. However,
our results suggest that LSBGs at both low and high masses can form
as a result of statistical variation in the processes
that also form HSBGs, namely the secular growth of galaxies from 
gas within their host halo at varying cosmic times, and the buildup
of the diffuse outer regions of galaxies due to mergers.
 
\section*{Acknowledgements}

The authors would like to thank Claudia Lagos
for providing data that was used in an early draft of this paper.
AK is grateful to Joop Schaye for helpful discussions.

AK, GG, and NP acknowledge support from CONICYT project Basal AFB-170002. GG and
NP were supported by Fondecyt 1150300. NP acknowledges support from Fondecyt
1191813. This project has received funding from the European Union's Horizon 2020
Research and Innovation Programme under the Marie Sk\l{}odowska-Curie grant
agreement No. 734374.

This work used the DiRAC Data Centric system at Durham University, operated by the
Institute for Computational Cosmology on behalf of the STFC DiRAC HPC Facility
(www.dirac.ac.uk). This equipment was funded by BIS National E-infrastructure
capital grant ST/K00042X/1, STFC capital grant ST/H008519/1, and STFC DiRAC
Operations grant ST/K003267/1 and Durham University. DiRAC is part of the National
E-Infrastructure.

\bibliographystyle{mn2e}

\appendix
\section{Impact of Dust on Surface Brightness}
\label{appendix}

All B-band surface brightnesses presented in the main body of this paper
were computed using dust-free luminosities for the EAGLE stellar particles.
In this appendix, we estimate the effect of dust
on these surface brightness values using images produced with the Monte Carlo
radiative transfer code SKIRT \citep{skirt2003, skirt2011, skirt2015}.
Only a full radiative transfer analysis can properly 
take into account the relative distribution of stars and dust in
each galaxy when estimating the dust obscuration. 

We use the 60 pkpc per side
$u$ and $g$ band images created for the EAGLE catalog for
galaxies with $M_{*} > 10^{10} \mathrm{M_{\odot}}$, and also supplement
these with newly created images for galaxies with lower masses
and $R_{28B} > 30$ pkpc. The full details of the SKIRT modeling
used for both the EAGLE catalog and our new images can be found
in \citet{trayford2017}. Here we provide a brief summary.

The inputs to SKIRT for each galaxy are 
the spatial distribution of emitted stellar light
and the dust distribution. The former is computed similarly
as in \citet{trayford2015}. Each stellar particle older than
100 Myr is assumed to be described by a simple stellar population,
for which the initial particle mass, stellar age, and SPH smoothed
metallicity are used to compute a \textsc{galaxev} \citet{bruzualcharlot} 
spectral energy distribution (SED), assuming a \citet{chabrier} initial mass function. 
The mass resolution of EAGLE is such that young stellar populations
are poorly sampled. Thus, star-forming gas particles and stellar particles younger than 100 Myr
are ``re-sampled'': their associated stellar mass is 
subdivided into stellar populations
with mass distribution based on the observed mass function
of Milky Way molecular clouds \citep{heyer2001}, and each subpopulation is
stochastically assigned a formation time based on the star formation
rate of the parent gas particle. Subpopulations older
than 10 Myr are assigned \textsc{galaxev} SEDs. Stellar
populations younger than 10 Myr are assigned \textsc{mappings-iii} \citep{groves2008}
SEDs, which include a prescription for dust absorption within (unresolved)
stellar birth clouds based on \citet{jonsson2010}.
Spatial smoothing is then applied to the star particles to produce a continuous spatial
luminosity distribution for the galaxy.

The dust spatial distribution is computed assuming that dust
traces the distribution of cold, metal-rich gas.
The dust-to-metal ratio is taken to be constant:
\begin{equation}
f_{\mathrm{dust}} = \frac{\rho_{\mathrm{dust}}}{Z\rho_{\mathrm{gas}}} = 0.3,
\end{equation}
where $\rho_{\mathrm{gas}}$ is the density of
gas that is either star forming or has $T < 8000$ K, and
$Z$ is its metallicity.
This value of $f_{\mathrm{dust}}$ was chosen such that applying it
to the population of EAGLE galaxies reproduces FIR observations \citep{camps2016}.
The dust model of \citet{zubko2004} is assumed,
and the dust optical depth is computed over an adaptively defined (AMR) grid.

Using these inputs, SKIRT produces a mock IFU image of 
each galaxy, assuming a given galaxy orientation and distance from the detector.
We utilize the $u$ and $g$ band images originally made
by \citet{trayford2017}, which are 60 pkpc on a side 
with resolution (pixel size) of 0.23 kpc, and were created
for all galaxies with $M_{*} > 10^{10} \mathrm{M_{\odot}}$. We convert
these images to the B band using the same equation from Lupton (2005)
as in the body of the paper (Eqn. \ref{lupton}).
We have also produced the following additional images in the B-band:
\begin{itemize}
\item images of galaxies with $10^{9.5} < M_{*}/\mathrm{M_{\odot}} < 10^{10}$, 
with size 90 pkpc per side in order to encompass $R_{28B}$ for all galaxies at these masses;
\item images with size 300 pkpc per side, for those galaxies with $M_{*} > 10^{10} \mathrm{M_{\odot}}$
and $R_{28B} \gtrsim 30$ pkpc. 
\end{itemize}

Because radiative transfer is computationally expensive, we used a resolution for these images
lower than that of the EAGLE catalog images, with pixel size 0.94 pkpc.
To further decrease the processing time, we only created new images for
those galaxies with a substantial amount of dust, as other galaxies
are expected to be minimally affected by dust extinction.
We use the dust resolution criterion defined by \citet{camps2018}:
$N_{\mathrm{dust}} > 250$ in the EAGLE catalog, where $N_{\mathrm{dust}}$
is the number of ``dust particles'', defined as the larger of the
number of star forming gas particles and the number of cold gas particles.
\citet{camps2018} determined that galaxies following this criterion
generally have well-resolved spatial dust distributions that can be used
by SKIRT. Thus the results shown in this appendix are only
for those galaxies that contain significant amounts of dust.

In the body of the paper, we measure the surface brightnesses of all galaxies in the face-on
orientation, based on the eigenvectors of the moment-of-inertia tensor. We similarly
choose SKIRT images that are oriented ``face-on''; however, EAGLE catalog ``face-on'' SKIRT
images are oriented with the angular momentum of the galaxy's stellar component 
towards the line of sight. We thus recompute
our dust-free particle-based surface brightness values with
the galaxies oriented based on their angular momentum vector. For disk galaxies, the minor axis
and the direction of the spin vector are generally well aligned, but
some elliptical galaxies are prolate and rotate about their major axis \citep{thob2019}.
While the latter galaxies are also old and not expected to contain a significant amount
of dust, we nevertheless note that for a small minority of galaxies presented in this appendix,
 the impact of dust may be overestimated relative
to what it would be for the galaxy orientation presented in the body of the paper.

From the mock IFU image created by SKIRT, we can measure
$R_{28B}$ and $\langle\mu_{B}\rangle$. Measuring these parameters
from a mock image is inherently somewhat different than doing so
using discrete stellar particles. We adopt a method that is intended to
 be similar to the manner in which we obtain these parameters from the
particle data. We assume that the isophotes of the galaxy are ellipses
that have equal orientation and axis ratio throughout the image, as we do for the particles. These
are obtained by finding the direction of maximum variance in the image,
and taking the square root of the ratio of this variance to the variance in the perpendicular
direction. We then compute the mean local surface brightness in elliptical annuli with this orientation and axis ratio
and find the one that has a value of 28 mag/asec$^{2}$, taking that smallest such
annulus if there is more than one. This determines $R_{28B}$. We then simply
compute the mean surface brightness within $R_{28B}$ to obtain $\langle\mu_{B}\rangle$.

By nature, $\langle\mu_{B}\rangle$ is highly dependent on $R_{28B}$ for a galaxy
of fixed total luminosity. The finite spatial resolution of SKIRT images imposes some
``smoothing'' on the surface brightness profiles derived from them, as does
the kernel smoothing of individual particles represented by Monte Carlo sampling. This means that,
while one would expect $R_{28B}$ to be the same or smaller when computed with dust
extinction than without, this is sometimes not the case, and $R_{28B}$
measured from the SKIRT image can be somewhat larger than 
that measured from the non-extincted star particles. 

We are able to quantify the effect of
 differing SKIRT image resolutions by using a 
subsample of galaxies with $R_{28B} \approx 30$ pkpc whose radii we measure
in both the original EAGLE images (with resolution 0.23 pkpc)
and in the larger 300 pkpc B-band images we have created that have a resolution
of 0.94 pkpc. We find that $R_{28B}$ measured from the latter images
is on average $4\%$ larger than that measured from the former, and $\langle\mu_{B}\rangle$ is
correspondingly 0.08 mag/asec$^{2}$ larger. Additionally,
the one sigma scatter between the values of $R_{28B}$ and $\langle\mu_{B}\rangle$ measured
at the two resolutions is $4\%$ and $\pm0.08$ mag/asec$^{2}$, respectively.
Given that the size of the uncertainty on $\langle\mu_{B}\rangle$ measured 
from the star particles is 0.1 mag/asec$^{2}$ (\S\ref{ressb}),
the difference in resolution between the two subsets of SKIRT images used
in this appendix should not have a significant effect our conclusions.

\begin{figure}
  \begin{center}
    \includegraphics[width=\columnwidth]{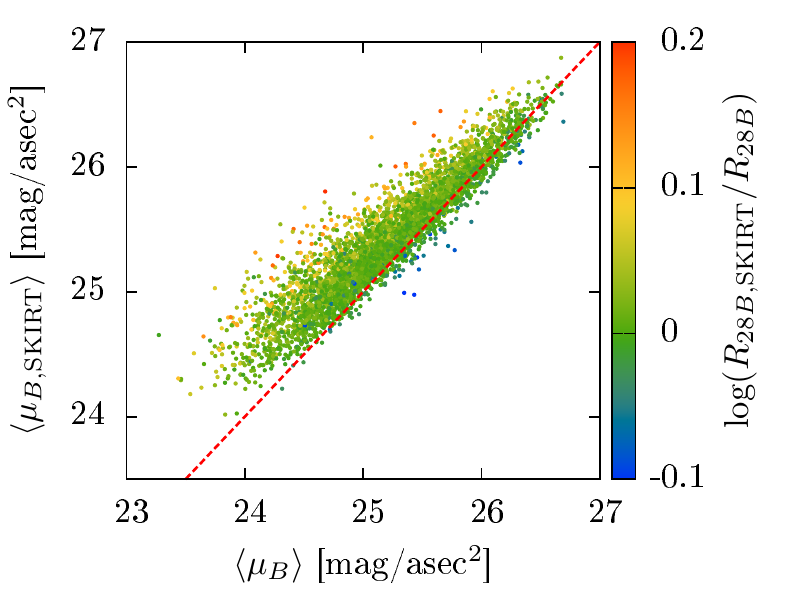}
    \caption{$\langle\mu_{B}\rangle$, the dust-free surface brightness within the 28 mag/asec$^{2}$
B-band isophote computed from the EAGLE stellar particles,
versus $\langle\mu_{B, \mathrm{SKIRT}}\rangle$, the equivalently-defined surface brightness
measured from the SKIRT images including dust extinction.
The only galaxies shown from our sample are those that contain sufficient dust
to accurately resolve its distribution (see text for criteria). The red dashed line
marks the line of equality. Color coding represents the ratio of the 28 mag/asec$^{2}$
B-band isophotal radius measured from the SKIRT image to that measured from the star particles.} 
    \label{app1}
  \end{center}
\end{figure}

\begin{figure}
  \begin{center}
    \includegraphics[width=\columnwidth]{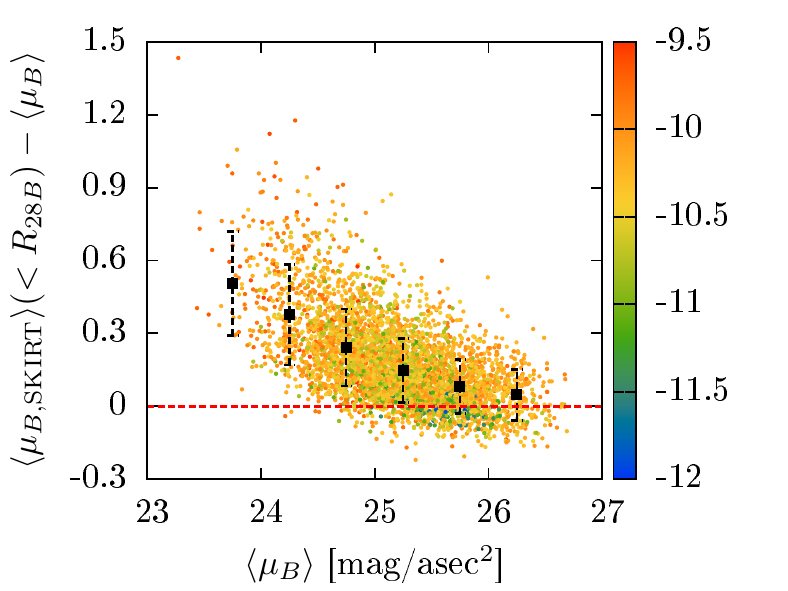}
    \caption{Similar to Figure \ref{app1}, but
now showing $\langle\mu_{B, \mathrm{SKIRT}}\rangle$ measured
within the same size radial aperture ($R_{28B}$) as was used to
measure $\langle\mu_{B}\rangle$ from the EAGLE star particles.
The shading of each bin represents the log-number of points within;
bins with only one point are represented by the individual point.
We note that the stochasticity inherent to Monte Carlo
radiative transfer means that the surface brightness derived
from SKIRT will sometimes be slightly brighter than that derived
from the EAGLE particles for galaxies whose dust extinction
is close to zero.}
    \label{app2}
  \end{center}
\end{figure}

4415 of 6987 galaxies in our sample contain sufficient dust that
we measure their $R_{28B}$ and $\langle\mu_{B}\rangle$ in dust-extincted
SKIRT images. In Figure \ref{app1}, we show $\langle\mu_{B}\rangle$ as measured
from the dust-free EAGLE particles versus $\langle\mu_{B, \mathrm{SKIRT}}\rangle$
measured from the SKIRT image including dust extinction. The color coding
shows the ratio of $R_{28B}$ measured from the SKIRT image 
($R_{28B, \mathrm{SKIRT}}$) to that
measured from the EAGLE particles. $\langle\mu_{B, \mathrm{SKIRT}}\rangle$
is on average 0.22 mag/asec$^{2}$ fainter than $\langle\mu_{B}\rangle$ 
from the particles. However, we see that there is a trend such 
that bright galaxies are more strongly affected by dust extinction than
faint ones. The effect is nevertheless not strong enough 
to greatly change the relative ordering of galaxies by surface brightness.
There are a small number of visible outliers for which $\langle\mu_{B}\rangle$ 
from SKIRT is much fainter than that from the particles. Visual
inspection of the SKIRT images reveals that these galaxies
generally have large amounts of dust extinction (visible as
dust lanes) in their bright nuclear regions. This decreases
the mean surface brightness within $R_{28B}$.  

We can also see in Figure \ref{app1} that there is a correlation between
the difference between $\langle\mu_{B, \mathrm{SKIRT}}\rangle - \langle\mu_{B}\rangle$ 
and $R_{28B, \mathrm{SKIRT}}/R_{28B}$. For the majority
of galaxies, the value of $R_{28B}$ measured from the SKIRT
images and the star particle distribution is similar. Extinction
in the outer parts of the galaxy can cause $R_{28B}$ to be smaller
in the SKIRT images, making the galaxy appear more compact
and its $\langle\mu_{B, \mathrm{SKIRT}}\rangle$ brighter.
However, cases in which $R_{28B}$ in the SKIRT images is larger
than that measured from the particles are likely the result of 
the finite resolution and shot noise in the SKIRT images.

To ensure that our estimate of the dust extinction
is not highly biased by these effects, we also compute
$\langle\mu_{B, \mathrm{SKIRT}}\rangle$ within the same
$R_{28B}$ used to measure $\langle\mu_{B}\rangle$
from the particles. This is shown in Figure \ref{app2}.
The distribution of $\langle\mu_{B, \mathrm{SKIRT}}\rangle$ in
this figure is similar to that in Figure \ref{app1},
implying that the majority of the effect seen there
is due to dust extinction within $R_{28B}$ rather than
the measured value of the latter. The 
galaxies in Figure \ref{app2} are on average 
0.18 mag/asec$^{2}$ fainter with dust extinction
than without; however, this value decreases from 
0.51 mag/asec$^{2}$ for galaxies with (dust-free) 
$23.5 < \langle\mu_{B}\rangle < 24$ to 0.05 mag/asec$^{2}$
for galaxies with $26 < \langle\mu_{B}\rangle < 26.5$.

We repeat that the galaxies shown in Figures \ref{app1} and \ref{app2} are only
those that contain a substantial amount of dust. 
In particular, all galaxies shown in these figures have non-zero
star formation rates, whereas many of the galaxies in our
sample do not (Figure \ref{fig3}); this is due to the 
assumptions made about the dust distribution following
that of the star-forming/cold gas. Dust thus has
the effect of slightly lowering the influence of SFR on the 
surface brightness (seen in Figure \ref{fig3}).
However, the difference is significantly smaller than that caused
by the SFR itself: the brightest galaxies are made
fainter by $\approx 0.5$ mag/asec$^{2}$, whereas their SFR increases
their brightness relative to a passive galaxy by $\approx 1.5$ mag/asec$^{2}$.

The Spearman correlation coefficient between
$\langle\mu_{B}\rangle$ and $\langle\mu_{B, \mathrm{SKIRT}}\rangle$
for galaxies with substantial dust is 0.93
with separately measured $R_{28B}$ and 0.96
within the same $R_{28B}$. If we assume galaxies without
substantial dust have $\langle\mu_{B}\rangle$ equal
to $\langle\mu_{B, \mathrm{SKIRT}}\rangle$ 
(which we find to be approximately
true for those dust-free galaxies that have EAGLE $u$ and $g$ SKIRT images),
these correlation coefficients remain unchanged.
Overall, the effect of dust on the surface
brightness is small for the majority of galaxies
and does not significantly affect the relative ordering of galaxies
in terms of their surface brightness.
This is partly because we take all galaxies to be face-on,
which is the orientation that minimizes the dust extinction. 
We thus assume that it is justified to ignore the
effect of dust in the body of this paper.

\label{lastpage}

\end{document}